\documentclass[onecolumn]{IEEEtran}
\usepackage{amssymb,amsmath,amsfonts,amsthm}
\usepackage{color}
\usepackage{cases}
\usepackage{standalone}
\usepackage{mathtools}
\usepackage{float}
\usepackage{cite}
\usepackage{suffix}
\usepackage{dblfloatfix}
\restylefloat{table}
\usepackage{amsfonts}
\usepackage{booktabs}
\usepackage{siunitx}
\usepackage{mathrsfs}
\usepackage{relsize}
\usepackage{graphicx}		
\usepackage{epstopdf}	
\usepackage{pgfplots}
\usepackage{pgfplotstable}
\usepackage{tikzscale} 
\usepackage{tikz}
\usepackage{subcaption}
\usepackage[font=small]{caption}
\usetikzlibrary{shapes}
\usepackage{algorithm}
\pgfplotsset{compat=newest}	
\pgfplotsset{plot coordinates/math parser=false}
\newlength\figureheight 
\newlength\figurewidth
\usepackage{tabularx}
\usepackage{enumerate}
\usepackage{mathrsfs}
\usepackage{amsbsy}
\usepackage{units}
\usepackage{bm}

\newcommand*\diff{\mathop{}\mathrm{d}}
\usepackage{textcomp}
\usepackage{placeins}
\usepackage{bbm}
\DeclareMathAlphabet{\mathpzc}{OT1}{pzc}{m}{it}
\usepackage{textgreek}
\newcommand{\ROME}[1]{%
\textup{\uppercase\expandafter{\romannumeral#1}}%
}
\usepackage{setspace}
\DeclarePairedDelimiterX\MeijerM[3]{\lparen}{\rparen}%
{\begin{smallmatrix}#1 \\ #2\end{smallmatrix}\delimsize\vert\,#3}
\newcommand\MeijerG[8][]{%
\mathcal{G}^{\,#2,#3}_{#4,#5}\MeijerM[#1]{#6}{#7}{#8}}
\WithSuffix\newcommand\MeijerG*[7]{%
\mathcal{G}^{\,#1,#2}_{#3,#4}\MeijerM*{#5}{#6}{#7}}
\theoremstyle{remark}
\newtheorem{thm}{Theorem}

\newtheorem{apx}{Approximation}

\makeatletter
\let\MYcaption\@makecaption
\makeatother
\usepackage[font=footnotesize]{subcaption}
\makeatletter
\let\@makecaption\MYcaption
\makeatother
\allowdisplaybreaks[4]
  
\begin{document}

\title{Full-Duplex Cloud Radio Access Network: Stochastic Design and Analysis} 
\author{Arman Shojaeifard, \textit{Member,~IEEE}, Kai-Kit Wong, \textit{Fellow,~IEEE},\linebreak Wei Yu, \textit{Fellow,~IEEE},  Gan Zheng, \textit{Senior Member,~IEEE},\linebreak and Jie Tang, \textit{Member,~IEEE}
}

\pagestyle{empty}

\maketitle

\begin{abstract} 

Full-duplex (FD) has emerged as a disruptive communications paradigm for enhancing the achievable
spectral efficiency (SE), thanks to the recent major breakthroughs in self-interference (SI) mitigation. The FD versus half-duplex (HD) SE gain, in cellular networks, is however largely limited by the mutual-interference (MI) between the downlink (DL) and the uplink (UL). A potential remedy for tackling the MI bottleneck is through cooperative communications. This paper provides a stochastic design and analysis of FD enabled cloud radio access network (C-RAN) under the Poisson point process (PPP)-based abstraction model of multi-antenna radio units (RUs) and user equipments (UEs). We consider different disjoint and user-centric approaches towards the formation of finite clusters in the C-RAN. Contrary to most existing studies, we explicitly take into consideration non-isotropic fading channel conditions and finite-capacity fronthaul links. Accordingly, upper-bound expressions for the C-RAN DL and UL SEs, involving the statistics of all intended and interfering signals, are derived. The performance of the FD C-RAN is investigated through the proposed theoretical framework and Monte-Carlo (MC) simulations. The results indicate that significant FD versus HD C-RAN SE gains can be achieved, particularly in the presence of sufficient-capacity fronthaul links and advanced interference cancellation \nolinebreak[4] capabilities.

\end{abstract}


\section{Introduction}

Full-duplex (FD) communications, that is simultaneous transmission and reception of wireless signals, has emerged as a disruptive solution for enhancing the achievable spectral efficiency (SE) \cite{Choi2010,khandani2010,5757799}. In the past, operating in FD mode was deemed unfeasible, due to the overwhelming self-interference (SI) which arises from the bi-directional wireless functionality. In recent years, significant technological advances have been made towards tackling the SI directly in FD mode, using any combination of passive suppression and active cancellation in analog and/or digital domains, see, e.g., \cite{5985554,6353396,6736751,7105651}. In point of fact, several protocols and prototypes for FD radios have been successfully implemented in practice, achieving near two-fold increase in SE versus the conventional half-duplex (HD) radios \cite{2011arXiv1107.0607S,Jain2011,Bharadia2013}. On the other hand, it has been shown that the large-scale FD functionality, in the context of cellular networks, is largely limited by the mutual-interference (MI) between the downlink (DL) and the uplink (UL) \cite{7105650,7736037,ArmanFDMM}. A potential remedy for tackling the MI bottleneck, and hence unlocking the end-to-end benefits of FD operation in cellular networks, may be through cooperative communications. 

Cloud radio access network (C-RAN) is a novel cellular network architecture in which the base station (BS) baseband processing and radio-frequency functionalities are decoupled \cite{guan2010,mobile2011c}. C-RAN facilitates cooperative wireless communications on a large-scale basis \cite{WYcloud}, with central processors (CPs) handling the baseband processing and, with the aid of fronthaul links, exchanging information with distributed radio units (RUs), which in turn, provide (for the most part) radio-frequency functionalities. C-RAN has received a great deal of attention in recent years thanks to its ability to address the inter-cell interference phenomenon, and in turn, allowing for higher SE and energy efficiency (EE) performance to be achieved \cite{7000980,6897914,7096298}. In addition, the use of cloud-computing-powered CPs and small-sized low-power RUs is shown to result in significant improvements in terms of deployment cost versus the conventional long-term-evolution (LTE) networks \cite{7018201,7110548}. At the present time, a particular attention is placed on content caching strategies in C-RAN such to further improve the underlying performance measures, such as delay, backhauling, and quality-of-experience (QoE), see, e.g., \cite{7488289,7880663}. 

A fundamental question hence arises, namely, what is the underlying FD versus HD SE gain in the context of cooperative wireless communications systems, and in particular, C-RAN? This paper takes a step in this direction by providing a stochastic design and analysis of such systems.

\subsection{Related Works}

The performance of C-RAN has been analyzed in the literature. In \cite{6600939}, the C-RAN outage probability in the DL was characterized considering a Poisson point process (PPP)-based RU deployment, and the minimum spatial density of RUs required for meeting a target SE was studied. Analytical expressions for the C-RAN outage probability and throughput in the DL were derived considering Mat\'{e}rn Hard-Core point process (MHCPP)-based RU deployment in \cite{7415954}. In particular, different RU selection schemes, under linear zero-forcing (ZF) precoding, were modeled and compared with one another. Other performance metrics for C-RAN, namely, physical (PHY)-layer security and EE, in the DL, were studied using the PPP-based abstraction model of BSs and RUs in \cite{7544500}. In particular, it was shown that the integration of massive multiple-input multiple-output (MIMO) assisted macro-cells with C-RAN can greatly improve the secrecy capacity and network EE. Recently in \cite{7738565}, the authors derived explicit expressions for the coverage and rate in DL C-RAN with finite clustering and limited channel knowledge. In particular, the promising potential of C-RAN, even in the presence of finite cooperative clusters and partial feedback, was further confirmed. It is also important to highlight the earlier information-theoretic works on cooperative wireless communications such as \cite{1678166,5594708,6015601,simeone2012,6482234,6376184}.

Several studies on the performance of FD enabled cooperative wireless systems have also been reported in the literature. In \cite{6845310}, an information-theoretic analysis of C-RAN with FD RUs based on the classical Wyner cellular model was provided. In particular, the authors investigated the FD C-RAN DL and UL SEs (versus single-cell processing), considering capacity-limited fronthaul links, successive interference cancellation (SIC) capability at the user equipment (UE) side, and perfect SI cancellation capability at the RU side. In addition, in \cite{7218556}, the authors considered a FD enabled multi-cell network MIMO paradigm, and utilized spatial interference-alignment (IA) towards tackling the MI from the UL operation on the DL performance. In particular, the scaling multiplexing gain of FD versus HD operation in multi-cell network MIMO was characterized in closed-form. On the other hand, the authors in \cite{2016arXiv160208836M}, considered a C-RAN scenario in which a single FD UE simultaneously communicates with randomly-deployed HD multi-antenna RUs in the DL and UL directions.  The results indicated that with appropriate beamforming and RU association, significant FD versus HD SE gains can be achieved for this particular case, subject to residual SI power being low. 

\subsection{Contributions}

In this work, we aim to investigate the potential SE gain of FD versus HD operation in the context of C-RAN. To this end, we provide a stochastic design analysis of large-scale cooperative cellular networks using the PPP-based abstraction model of multi-antenna BSs and UEs. To the best of our knowledge, this is the first work that, with the aid of stochastic geometry theory, studies the C-RAN SE performance with FD RUs (relays), equipped with multiple transmit and receive antennas. Apart from the immediately apparent new challenges posed by the FD operation in our setup, our proposed framework differs from the theoretical models (for HD C-RAN) in the existing literature on multiple fronts, as is highlighted below, and throughout this \nolinebreak[4] paper. 

Here, we consider different disjoint and user-centric approaches towards the formation of finite clusters in the C-RAN. In the former, the C-RAN comprises fixed non-overlapping clusters, whereas in the latter, (potentially overlapping) clusters are formed around every scheduled UE, respectively. Within any finite cluster, a CP is considered to facilitate cooperative communications between the RUs and UEs. To our knowledge, this work, as well as the recent contribution in \cite{7738565} for HD C-RAN, may be viewed as the only stochastic geometry-based models for C-RAN which take into account the impact of inter-cluster interference due to finite clustering. Many related works, on the other hand, consider perfect coordination across all clusters, which is not feasible due to the practical constraints (such as propagation delay). 

In this work, in contrast to the existing theoretical studies for C-RAN (e.g., \cite{6600939,7415954,7544500,7738565,6845310,2016arXiv160208836M}), we explicitly take into consideration the non-isotropic nature of wireless channels, which inherently arises as a result of cooperative communications. For example, in the context of C-RAN, in each cluster, the channels between the multi-antenna RUs and UEs (which follow from independent PPPs) involve different distance-dependent path-loss parameters. Hence, with cooperative beamforming, the intended and interfering channels, involve non-identically distributed elements. Here, building on the results from \cite{6881264,6823643,7536884,7450690}, we utilize the Gamma moment-matching technique in order to characterize the distributions of all intended and interfering signals with cooperative ZF beamforming in the FD multi-cluster multi-user C-RAN under consideration.  
 
Cooperative beamforming and resource allocation problems under different fronthaul strategies and constraints in C-RAN have been extensively studied in the optimization-related literature (see, e.g., \cite{WYcloud,7590010,7809154}). On the other hand, the consideration of finite fronthaul capacity, for the most part, is missing from the existing theoretical C-RAN models (besides the work in \cite{6845310}, for a Wyner-based topology). In this work, we incorporate the impact of capacity-limited fronthaul links in the proposed theoretical framework using cut-set bounds on the achievable capacity \cite{Cover1991}. Accordingly, we derive upper-bound expressions for the FD C-RAN DL and UL SEs, in particular, as a main technical contribution of this work, fully accounting for the MI in the DL (i.e., UE-UE interference) and the UL (i.e., BS-BS interference). 

The validity of the theoretical findings is confirmed through Monte-Carlo (MC) simulations. The results highlight the promising potential of FD versus HD operation with regards to the C-RAN SE performance, particularly, in the presence of sufficient-capacity fronthaul links and advanced interference cancellation capabilities. Our results further confirm that the underlying SE gains in FD versus HD C-RAN, compared to that in conventional cellular systems, can be significantly higher as a result of cooperative wireless communications. 

\subsection{Organization}

The remainder of this paper is organized as follows. In Section \ROME{2}, the C-RAN model and operation under consideration is described. We derive the distributions of all intended and interfering non-isotropic channel power gains in Section \ROME{3}. The theoretical analysis of C-RAN SE is given in Section \ROME{4}. Numerical results are then provided in Section \ROME{5}. Finally, conclusions are drawn in Section \ROME{6}.

\subsection{Notation}

The following notation is used throughout this paper. $\boldsymbol{X}$ is a matrix with ($i,j$)-th entry $\{ \boldsymbol{X} \}_{i,j}$; $\boldsymbol{x}$ is a vector with $k$-th element $\{ \boldsymbol{x} \}_{k}$; $T$, $\dag$, and $+$ are the transpose, Hermitian, and pseudo-inverse operations; $j$ is the imaginary unit; $\operatorname{Im} (.)$ is the imaginary part; $\mathbb{E}_{x}\{.\}$ is the expectation; $\mathscr{P} (.)$ is the probability; $\mathcal{F}_{x}(.)$ is the cumulative distribution function (CDF); $\mathcal{P}_{x}(.)$ is the probability density function (PDF); $\mathcal{M}_{x}(.)$ is the moment-generating-function (MGF); $| x |$ is the modulus; $\| \boldsymbol{x} \|$ is the Euclidean norm; $\mathbf{I}_{(.)}$ is the identity matrix; $\text{Null}(.)$ is a nullspace; $\mathcal{C}\mathcal{N}(\mu,\nu^2)$ is the circularly-symmetric complex Gaussian distribution with mean $\mu$ and variance $\nu^{2}$; $\Gamma(.)$ and $\Gamma(.,.)$ are the Gamma and incomplete (upper) Gamma functions; $\mathpzc{G} (\kappa,\theta)$ is the Gamma distribution with shape parameter $\kappa$ and scale parameter $\theta$; and $_2F_1(.,.;.;.)$ is the Gauss hypergeometric function, respectively.

\section{System Description}

\subsection{Network Topology}

In this work, we consider a multi-cluster multi-user C-RAN in which the RUs and UEs are deployed on the two-dimensional (2D) Euclidean space according to independent stationary PPPs $\Phi^{(d)}$ and $\Phi^{(u)}$ with spatial densities $\lambda^{(d)}$ and $\lambda^{(u)}$, respectively.
In FD C-RAN, each FD RU, equipped with $N^{(d)}$ transmit and $N^{(u)}$ receive antennas, is considered to be serving $\mathcal{K}^{(d)}$ ($\leq N^{(d)}$) HD DL UEs and $\mathcal{K}^{(u)}$ ($\leq N^{(u)}$) HD UL UEs, all equipped with single antennas, per resource block. On the other hand, in HD C-RAN, the DL and UL occur over different resource blocks. In what follows, we provide a description for the FD C-RAN. With apparent adjustments, the HD C-RAN model and operation can be depicted. The impact of post-processing (residual) SI, in line with the findings in the related literature \cite{6845310,ArmanFDMM,AtzeniK16a}, is considered to be negligible compared to the MI. It should be noted the assumption of HD UEs is made due to the inherent restrictions of legacy devices, and the high cost of equipping FD functionality at the UEs, at least in the foreseeable future \cite{7891794}. Otherwise, the proposed framework can be readily modified to account for FD enabled UEs. 

\subsection{Finite Clustering}

Here, we consider different disjoint and user-centric approaches for the formation of the finite clusters in the C-RAN. In the former, the C-RAN comprises disjoint non-overlapping clusters, whereas in the latter, clusters are formed around each scheduled UE. In the case of disjoint clustering, a typical UE is uniformly located in the corresponding cluster area. In this sense, the performances of the different UEs, for example one that is located at the cluster center, versus one located at the cluster edge, are inherently different. The user-centric clustering approach can be viewed as a remedy for tackling the poor performance at the cell-edges. However, the different clusters may be overlapping under the user-centric approach, hence, giving rise to intra-cluster interference. Within any finite cluster, a CP is considered to facilitate cooperative communications between the multi-antenna RUs and UEs. 

\subsection{Channel Model}

We proceed by defining the different channels. Note that the letters ``$g,\boldsymbol{g},\boldsymbol{G}$" and ``$h,\boldsymbol{h},\boldsymbol{H}$" are accordingly used to distinguish between the effective DL and UL channels, respectively. Moreover, the letters ``$f,\boldsymbol{f},\boldsymbol{F}$" correspond to small-scale channel attenuation, whereas large-scale fading effects are represented using the letter $\beta$. In this work, we utilize the Rayleigh distribution to model the small-scale fading channels. Moreover, we employ the unbounded path-loss model with exponent $\alpha$ ($> 2$), i.e., $\beta_{a,b} = r^{- \alpha}_{a,b}$, where $r_{a,b}$ denotes the Euclidean distance between the nodes $a$ and $b$. Note that the number of cooperative RUs in a cluster $c$ is denoted with $L_{c}$. The transmit powers of the multi-antenna RUs and UEs are set as $p^{(d)}$ (per-user) and $p^{(u)}$, respectively. 

Let $\boldsymbol{g}_{m_{l},c_{k}} = \sqrt{\beta_{m_{l},c_{k}}} \boldsymbol{f}_{m_{l},c_{k}}$, where $\boldsymbol{f}^{T}_{m_{l},c_{k}} \sim \mathcal{C} \mathcal{N} ( \boldsymbol{0},\mathbf{I}_{N^{(d)}} )$, denote the DL channel from the RU $l$ in the cluster $m$ to the UE $k$ in the cluster $c$. The combined DL channel from the $L_{m}$ cooperative RUs in the cluster $m$ to the UE $k$ in the cluster $c$ is represented using $\boldsymbol{g}_{m,c_{k}} = [\boldsymbol{g}_{m_{l},c_{k}}]_{1 \leq l \leq L_{m}} \in \mathcal{C}^{1 \times L_{m} N^{(d)}}$. Moreover, we use $\boldsymbol{h}_{c_{k},m_{l}} = \sqrt{\beta_{c_{k},m_{l}}} \boldsymbol{f}_{c_{k},m_{l}}$, where $\boldsymbol{f}_{c_{k},m_{l}} \sim \mathcal{C} \mathcal{N} ( \boldsymbol{0},\mathbf{I}_{N^{(u)}} )$ to denote the UL channel from the UE $k$ in the cluster $c$ to the RU $l$ in the cluster $m$. The combined UL channels to the cooperative RUs in the cluster $m$ from the UE $k$ in the cluster $c$ is given by $\boldsymbol{h}_{c_{k},m} = [\boldsymbol{h}^{T}_{c_{k},m_{l}}]^{T}_{1 \leq l \leq L_{m}} \in \mathcal{C}^{L_{m} N^{(u)} \times 1}$. The cross-mode channel from the UL UE $k$ in the cluster $m$ to the DL UE $o$ in the cluster $c$ is denoted with $h_{m_{k},c_{o}} = \sqrt{\beta_{m_{k},c_{o}}} f_{m_{k},c_{o}}$, where $f_{m_{k},c_{o}} \sim \mathcal{C} \mathcal{N} (0 , 1)$. On the other hand, the cross-mode channel from the RU $l$ in the cluster $m$ to the RU $b$ in the cluster $c$ is given by $\boldsymbol{G}_{m_{l},c_{b}} = \sqrt{\beta_{m_{l},c_{b}}} \boldsymbol{F}_{m_{l},c_{b}}$, where $\boldsymbol{F}_{m_{l},c_{b}} \sim \mathcal{C} \mathcal{N} ( \boldsymbol{0} ,\mathbf{I}_{N^{(u)} \times N^{(d)}} )$. We can then express the channel from the RU $l$ in the cluster $m$ to the cooperative RUs in the cluster $c$ using $\boldsymbol{G}_{m_{l},c} = [\boldsymbol{G}^{T}_{m_{l},c_{b}}]^{T}_{0 \leq b \leq L_{c}} \in \mathcal{C}^{L_{c} N^{(u)} \times N^{(d)}}$. In addition, we denote the channel from the cooperating RUs in the cluster $m$ to the cooperating RUs in the cluster $c$ using $\boldsymbol{G}_{m,c} = [\boldsymbol{G}_{m_{l},c}]_{1 \leq l \leq L_{m}} \in \mathcal{C}^{L_{c} N^{(u)} \times L_{m} N^{(d)}}$.  

\subsection{Baseband Signals}

Let $\boldsymbol{G}_{c} = [\boldsymbol{g}^{T}_{c,c_{k}}]^{T}_{1 \leq k \leq L_{c} \mathcal{K}^{(d)} } \in \mathcal{C}^{L_{c} \mathcal{K}^{(d)} \times L_{c} N^{(d)}}$ denote the combined DL channels from the cooperative RUs to the active DL UEs in the cluster $c$. Moreover, $\boldsymbol{s}_{c} = [s_{c,c_{k}}]^{T}_{1 \leq k \leq L_{c} \mathcal{K}^{(d)}} \in \mathcal{C}^{L_{c} \mathcal{K}^{(d)} \times 1}$, $\mathbb{E} \left\{ | s_{c,c_{k}} |^2 \right\} = 1$, denotes the DL complex symbol vector from the cooperative RUs to the active DL UEs in the cluster $c$. The normalized linear precoding matrix at the cluster $c$ is expressed as $\boldsymbol{V}_{c} = [\boldsymbol{v}_{c,c_{k}}]_{1 \leq k \leq L_{c} \mathcal{K}^{(d)}} \in \mathcal{C}^{L_{c} N_{t} \times L_{c} \mathcal{K}^{(d)}}$, $\mathbb{E} \{ \| \boldsymbol{v}_{c,c_{k}} \|^2 \} = 1$. The DL received signal for the reference DL active UE $c_{o}$ in the cluster $c$ can be represented as 
\begin{align}
y^{(d)} & = \underbrace{ \sqrt{p^{(d)}} \boldsymbol{g}_{c,c_{o}} \boldsymbol{v}_{c,c_{o}} s_{c,c_{o}}}_{\text{intended signal}} + \underbrace{\sqrt{p^{(d)}} \boldsymbol{g}_{c,c_{o}} \sum_{c_{k} \in \Psi^{(d)}_{c} \setminus \{ c_{o} \} } \boldsymbol{v}_{c,c_{k}} s_{c,c_{k}}}_{\text{intra-cluster interference}} + \underbrace{ \sqrt{p^{(d)}} \sum_{m \in \Psi \setminus \{ c \}} \boldsymbol{g}_{m,c_{o}} \boldsymbol{V}_{m} \boldsymbol{s}_{m} }_{\text{inter-cluster interference}} \nonumber \\ & + \underbrace{ \sqrt{p^{(u)}} \sum_{m \in \Psi,m_{k} \in \Psi^{(u)}_{m}} h_{m_{k},c_{o}} s_{m_{k},m} }_{\text{mutual-interference}} 
+ \underbrace{\eta^{(d)}}_{\text{noise}}
\end{align}
where $\Psi$ is the set of all clusters, $\Psi^{(d)}_{c}$ is the set of active DL UEs in the cluster $c$, $\Psi^{(u)}_{m}$ is the set of active UL UEs in the cluster $m$, $s_{m_{k},m}$ is the information symbol transmitted from active UL UE $k$ in cluster $m$, and $\eta^{(d)}$ is the zero-mean complex additive white Gaussian noise (AWGN) with variance $\nu^{(d)}$, respectively. 

On the other hand, let $\boldsymbol{H}_{c} = [\boldsymbol{h}_{c_{k},c}]_{1 \leq k \leq L_{c} \mathcal{K}^{(u)}} \in \mathcal{C}^{L_{c} N^{(u)} \times L_{c} \mathcal{K}^{(u)}}$ represent the collective UL channel from the active UL UEs at the cooperative RUs in the cluster $c$. The normalized linear decoding matrix at the CP in the cluster $c$ is given by $\boldsymbol{W}_{c} = [ \boldsymbol{w}^{T}_{c_{k},c} ]^{T}_{1 \leq k \leq L_{c} \mathcal{K}^{(u)}} \in \mathcal{C}^{L_{c} \mathcal{K}^{(u)} \times L_{c} N^{(u)}}$, $\mathbb{E} \{ \| \boldsymbol{w}_{c_{k},c} \|^{2} \} = 1$. The post-processing UL received signal from the reference active UL UE $c_{i}$ in the cluster $c$ is hence given by
\begin{align}
y^{(u)} & = \underbrace{\sqrt{p^{(u)}} \boldsymbol{w}^{T}_{c_{i},c} \boldsymbol{h}_{c_{i},c} s_{c_{i},c}}_{\text{intended signal}} + \underbrace{\sqrt{p^{(u)}} \sum_{c_{k} \in \Psi^{(u)}_{c} \setminus \{ c_{i} \} } \boldsymbol{w}^{T}_{c_{i},c} \boldsymbol{h}_{c_{k},c} s_{c_{k},c}}_{\text{intra-cluster interference}} + \underbrace{\sqrt{p^{(u)}} \sum_{m \in \Psi \setminus \{ c \},m_{k} \in \Psi^{(u)}_{m}} \boldsymbol{w}^{T}_{c_{i},c} \boldsymbol{h}_{m_{k},c} s_{m_{k},m} }_{\text{inter-cluster interference}} \nonumber \\ & + \underbrace{\sqrt{p^{(d)}} \sum_{m \in \Psi \setminus \{ c \} } \boldsymbol{w}^{T}_{c_{i},c} \boldsymbol{G}_{m,c} \boldsymbol{V}_{m} \boldsymbol{s}_{m} }_{\text{mutual-interference}}
 + \underbrace{\boldsymbol{w}^{T}_{c_{0},c} \boldsymbol{\eta}^{(u)}}_{\text{scaled noise}}
\end{align}
where $\boldsymbol{\eta}^{(u)} \in \mathcal{C}^{L_{c} N^{(u)} \times 1}$ is the circularly-symmetric zero-mean complex AWGN vector with covariance matrix $\nu^{(u)} \mathbf{I}_{L_{c} N^{(u)}}$. 

\subsection{Cooperative Beamforming}

In the DL, we adopt a cooperative ZF precoder for suppressing intra-cluster interference. The baseband processing is carried out at the CP in each cluster, and the corresponding information is forwarded via fronthaul links to the cooperative RUs. Specifically, in the cluster $c$, the cooperative ZF beamformer $\boldsymbol{V}_{c}$ is set equal to the normalized columns of $\boldsymbol{G}^{+}_{c} = \boldsymbol{G}^{\dag}_{c} ( \boldsymbol{G}_{c} \boldsymbol{G}^{\dag}_{c} )^{-1} \in \mathcal{C}^{L_{c} N^{(d)} \times L_{c} \mathcal{K}^{(d)}}$. Further, in the UL, the signals received at the cooperative RUs from the active UL UEs are compressed, and forwarded via fronthaul links to the CP. The CP, in turn, performs joint decoding. Here, we consider the case where the CP applies a cooperative ZF decoder for suppressing intra-cluster interference in the UL. Specifically, in the cluster $c$, the normalized rows of $\boldsymbol{H}^{+}_{c} = ( \boldsymbol{H}_{c}^{\dag} \boldsymbol{H}_{c} )^{-1} \boldsymbol{H}^{\dag}_{c} = [\boldsymbol{\hat{h}}^{T}_{c_{k},c}]^{T}_{1 \leq k \leq L_{c} \mathcal{K}^{(u)}} \in \mathcal{C}^{L_{c} \mathcal{K}^{(d)} \times L_{c} N^{(u)}}$ are set equal to the row vectors of $\boldsymbol{W}_{c}$.

\section{Signals Distributions} 

We proceed by defining the signal-to-interference-plus-noise ratios (SINRs) in the FD C-RAN under consideration.

The received SINR at the reference active DL UE $o$ in the cluster $c$ can be expressed as
\begin{align}
\gamma^{(d)} = \frac{\mathcal{X}^{(d)}}{\mathcal{ICI}^{(d)} + \mathcal{CMI}^{(d)} + \nu^{(d)}}
\end{align}
where $\mathcal{X}^{(d)} = p^{(d)} | \boldsymbol{g}_{c,c_{o}} \boldsymbol{v}_{c,c_{o}} |^{2}$, $\mathcal{ICI}^{(d)} = p^{(d)} \sum_{m \in \Psi} \| \boldsymbol{g}_{m,c_{o}} \boldsymbol{V}_{m} \|^2$, and $\mathcal{CMI}^{(d)} = p^{(u)} \sum_{m \in \Psi,m_{k} \in \Psi^{(u)}_{m}}$ $| h_{m_{k},c_{o}} |^2$.

On the other hand, in the cluster $c$, the received UL SINR for the reference active UL UE $i$ at the CP is given by
\begin{align}
\gamma^{(u)} = \frac{\mathcal{X}^{(u)}}{\mathcal{ICI}^{(u)} + \mathcal{CMI}^{(u)} 
 + \nu^{(u)}}
\end{align}
where $\mathcal{X}^{(u)} = p^{(u)} | \boldsymbol{w}^{T}_{c_{i},c} \boldsymbol{h}_{c_{i},c} |^{2}$, $\mathcal{ICI}^{(u)} = p^{(u)} \sum_{m \in \Psi \setminus \{ c \},m_{k} \in \Psi^{(u)}_{m}} | \boldsymbol{w}^{T}_{c_{i},c} \boldsymbol{h}_{m_{k},c} |^{2}$, and $\mathcal{CMI}^{(u)} = p^{(d)}$ $\sum_{m \in \Psi \setminus \{ c \}}$\linebreak $\| \boldsymbol{w}^{T}_{c_{i},c} \boldsymbol{G}_{m,c} \boldsymbol{V}_{m} \|^{2}$.

In the case of cooperative beamforming, for example in the C-RAN under consideration, the channels are non-isotropic in nature, given that the links between randomly-located RUs and active UEs involve different path-loss parameters. As a result, it is not possible to derive the exact distributions of the different intended and interfering channel power gains. It has been shown, e.g., in \cite{7450690}, that the Gamma moment matching technique can be invoked in order to derive approximate expressions for the intended and interfering channel power gains in the case of HD network MIMO. In what follows, we also apply the moment matching technique to characterize the different channel power gains in the context of FD C-RAN with finite clustering. Note that the average number of cooperating RUs per cluster is denoted with $L$. 

\begin{apx}
\label{apx1}
The DL and UL intended channel power gains with cooperative ZF beamforming in the FD C-RAN under consideration are respectively given by
\begin{align}
\left| \boldsymbol{g}_{c,c_{o}} \boldsymbol{v}_{c,c_{o}} \right|^{2} \approx \sum_{c_{j} \in \Phi^{(d)}_{c}} \beta_{c_{j},c_{o}} \psi_{c_{j},c_{o}}, \;\, \psi_{c_{j},c_{o}} \sim \mathpzc{G} \left(  N^{(d)} - \mathcal{K}^{(d)} + \frac{1}{L} , 1 \right)
\end{align}
\begin{align}
\left| \boldsymbol{w}^{T}_{c_{i},c} \boldsymbol{h}_{c_{i},c} \right|^{2} \approx \sum_{c_{l} \in \Phi^{(d)}_{c}} \beta_{c_{i},c_{l}} \psi_{c_{i},c_{l}}, \;\, \psi_{c_{i},c_{l}} \sim \mathpzc{G} \left(  N^{(u)} - \mathcal{K}^{(u)} + \frac{1}{L} , 1 \right).
\end{align}
\end{apx}

\begin{apx}
\label{apx2}
The DL and UL inter-cluster interference channel power gains with cooperative ZF beamforming in the FD C-RAN under consideration are respectively given by
\begin{align}
\left\| \boldsymbol{g}_{m,c_{o}} \boldsymbol{V}_{m} \right\|^2 \approx \sum_{m_{j} \in \Phi^{(d)}_{m}} \beta_{m_{j},c_{o}} \psi_{m_{j},c_{o}}, \;\, \psi_{m_{j},c_{o}} \sim \mathpzc{G} \left( \mathcal{K}^{(d)} , 1 \right) 
\end{align}
\begin{align}
\left| \boldsymbol{w}^{T}_{c_{i},c} \boldsymbol{h}_{m_{k},c} \right|^{2} \approx \sum_{c_{l} \in \Phi^{(d)}_{c}} \beta_{m_{k},c_{l}} \psi_{m_{k},c_{l}}, \;\, \psi_{m_{k},c_{l}} \sim \mathpzc{G} \left( \frac{1}{L} , 1 \right).  
\end{align}
\end{apx}

\begin{apx}
\label{apx3}
The DL and UL cross-mode interference channel power gains with cooperative ZF beamforming in the FD C-RAN under consideration are respectively given by
\begin{align}
\left| h_{m_{k},c_{o}} \right|^2 = \beta_{m_{k},c_{o}} \psi_{m_{k},c_{o}}, \;\, \psi_{m_{k},c_{o}} \sim \mathpzc{G} (1,1)
\label{eq:DLcmi}
\end{align}
\begin{align}
\left\| \boldsymbol{w}^{T}_{c_{i},c} \boldsymbol{G}_{m,c} \boldsymbol{V}_{m} \right\|^{2} \approx \sum_{m_{j} \in \Phi^{(d)}_{m}} \sum_{c_{l} \in \Phi^{(d)}_{c}} \beta_{m_{j},c_{l}} \psi_{m_{j},c_{l}}, \;\, \psi_{m_{j},c_{l}} \sim \mathpzc{G} \left( \frac{\mathcal{K}^{(d)}}{L} , 1 \right). 
\label{eq:ULcmi}
\end{align}
Proof: {\normalfont See Appendix \ref{apx:proof}.}
\end{apx}

\section{C-RAN Analysis}

\subsection{Unified Framework}

In practice, the C-RAN performance depends on the processing and relaying strategies under the finite-capacity fronthaul links. In general, the fronthaul constraint limits the information exchange between the CP and the cooperative RUs. The particular impact, however, significantly varies depending on the fronthaul technology (fiber optic or wireless), communications direction (DL versus UL), etc. For example, in the DL C-RAN, the CP may adopt data-sharing or compression-based relaying strategies. In the former, the fronthaul constraint restricts the cluster size (number of cooperative RUs), whereas in the latter, the capacity-limited fronthaul induces certain compression noise. Similarly, in the UL C-RAN, the impact of fronthaul largely depends on the infrastructure and choice of relaying strategy (e.g., compress-forward versus decode-forward) \cite{WYcloud}. 

In any case, in the C-RAN, the achievable relaying rate under finite-capacity fronthaul links is upper-bounded according to the information-theoretic cut-set theorem \cite{Cover1991}. In this work, we consider the case where the normalized (with respect the available bandwidth) per-user DL and UL capacities of the fronthaul links between the CP to each cooperative RU are $C^{(d)}$ and $C^{(u)}$ (in nat/s/Hz), respectively \cite{6845310}. Note that $\log \left( 1 + \gamma^{(d)} \right)$ and $\log \left(1 + \gamma^{(u)} \right)$ respectively denote the C-RAN instantaneous DL and UL per-user SEs (in nat/s/Hz). Hence, we can characterize the C-RAN DL and UL SE upper-bounds in the presence of capacity-limited fronthaul links. Note that $C^{(d)}, C^{(u)} \rightarrow + \infty$ corresponds to the case with ideal (infinite) capacity fronthaul.  
 
\begin{thm}
\label{thm1}
In the FD C-RAN under capacity-limited fronthaul links, the achievable per-user DL and UL SEs (in nat/s/Hz) can be respectively upper-bounded by the cut-set theorem as  	
\begin{align}
 \mathcal{S}^{(d)} \leq \mathbb{E} \left\{ \min \left( \log \left( 1 + \gamma^{(d)} \right) , C^{(d)} \right) \right\} = \int^{C^{(d)}}_{0} \frac{1}{1 + x} \left( 1 - \mathcal{F}_{\gamma^{(d)}} (x) \right) \diff x
\end{align}
\begin{align}
\mathcal{S}^{(u)} \leq \mathbb{E} \left\{ \min \left( \log \left( 1 + \gamma^{(u)} \right) , C^{(u)} \right) \right\} = \int^{C^{(u)}}_{0} \frac{1}{1 + x} \left( 1 - \mathcal{F}_{\gamma^{(u)}} (x) \right) \diff x. 
\end{align}
Proof: {\normalfont See Appendix \ref{thm1:proof}.}
\end{thm}

The FD C-RAN SE upper-bound expressions under the fronthaul constraint involve the CDFs of the SINRs. In the case of multi-antenna communications over isotropic Rayleigh fading channels, the coverage probability can be calculated in a number of ways, see, e.g., \cite{6775036,6596082,6862011} (the reader is referred to \cite{7812768} for multi-stream coverage performance analysis). On the other hand, no prior work has derived the SINR distributions in the case of cooperative multi-antenna communications with non-isotropic channels. In this work, we incorporate the Gil-Pelaez inversion theorem to derive, for the first time, explicit expressions for the FD C-RAN DL and UL coverage probabilities over non-isotropic fading channels. 

\begin{thm}
\label{thm2}
The CDFs of the DL and UL SINRs in the FD C-RAN under consideration are given by
\begin{align}
\mathcal{F}_{\gamma^{(d)}} (x) = \frac{1}{2} - \frac{1}{\pi} \int^{+ \infty}_{0} \frac{1}{s} \operatorname{Im} \left( \mathcal{M}_{\Theta^{(d)}} \left( j s \right) \exp \left( j s \nu^{(d)}  \right) \right) \diff s
\end{align}
\begin{align}
\mathcal{F}_{\gamma^{(u)}} (x) = \frac{1}{2} - \frac{1}{\pi} \int^{+ \infty}_{0} \frac{1}{s} \operatorname{Im} \left( \mathcal{M}_{\Theta^{(u)}} \left( j s \right) \exp \left( j s \nu^{(u)}  \right) \right) \diff s
\end{align}
where
\begin{align}
\mathcal{M}_{\Theta^{(d)}} \left( j s \right) = \mathcal{M}_{\mathcal{ICI}^{(d)}} \left( j s \right) \mathcal{M}_{\mathcal{CMI}^{(d)}} \left( j s \right) \mathcal{M}_{\mathcal{X}^{(d)}} \left( \frac{- j s}{x} \right)
\end{align}
\begin{align}
\mathcal{M}_{\Theta^{(u)}} \left( j s \right) = \mathcal{M}_{\mathcal{ICI}^{(u)}} \left( j s \right) \mathcal{M}_{\mathcal{CMI}^{(u)}} \left( j s \right) \left( j s \right) \mathcal{M}_{\mathcal{X}^{(u)}} \left( \frac{- j s}{x} \right). 
\end{align}
Proof: {\normalfont See Appendix \ref{thm2:proof}.}
\end{thm}

Note that in the case of infinite-capacity fronthaul links ($C^{(d)}, C^{(u)} \rightarrow + \infty$), the C-RAN DL and UL SEs can be obtained by utilizing the computationally-efficient non-direct MGF-based approach from \cite{5407601}, which avoids the need for the computation of the coverage probabilities. This method is highlighted below.  

\begin{thm}
\label{thm3}
In the FD C-RAN without constraint on the fronthaul capacity, the achievable per-user DL and UL SEs (in nat/s/Hz) can be respectively expressed as	
\begin{align}
\mathcal{S}^{(d)} = \int^{\infty}_{0} \left( 1 - \mathcal{M}_{\mathcal{X}^{(d)}} ( z ) \right) \mathcal{M}_{\mathcal{ICI}^{(d)}} ( z ) \mathcal{M}_{\mathcal{CMI}^{(d)}} ( z ) \frac{\exp \left(- z \nu^{(d)}\right)}{z} \diff z 
\end{align}
\begin{align}
\mathcal{S}^{(u)} = \int^{\infty}_{0} \left( 1 - \mathcal{M}_{\mathcal{X}^{(u)}} ( z ) \right) \mathcal{M}_{\mathcal{ICI}^{(u)}} ( z ) \mathcal{M}_{\mathcal{CMI}^{(u)}} ( z ) 
\frac{\exp \left(- z \nu^{(u)}\right)}{z} \diff z.
\end{align}
\end{thm}

\subsection{Disjoint vs. User-Centric Clustering}

Now we are ready to provide explicit expressions for the FD C-RAN DL and UL per-user SEs under different disjoint and user-centric clustering approaches. Note that the parameters with superscripts ``d/j" and ``u/c" correspond to the former and latter cases, respectively. Here, we approximate the disjoint (hexagonal) cluster by a circular region of radius $\mathcal{R}$ with the PDF of the arbitrary distance $r$ ($\geq 0$) of the reference user to the cluster center given by $\mathcal{P}_{r} (d) = \frac{2 d}{\mathcal{R}^{2}}$. In the case of user-centric clustering, each scheduled UE is located at the cluster center.  

In regards to the FD operation, we characterize the DL MI considering the UEs may be capable of performing SIC. In order to capture performance for general cases, we consider an exclusion region of radius $\mathcal{E}$ when modeling the UE-UE interference. Some special cases include (i) $\mathcal{E} = 0$, a worst-case scenario, without any interference cancellation capability, and (ii) $\mathcal{E} = \mathcal{R}$, a best-case scenario, where the UL intra-cluster signals are successively decoded and suppressed prior to the processing of the DL intended signals. Further, the UL MI (i.e., BS-BS interference) is characterized considering the sum interference from every inter-cluster RU with respect to each intra-cluster RU, all located randomly accordingly to the PPP-based abstraction model.

We proceed by defining the following functions which are subsequently used in the analysis 
\begin{align}
\Xi (y,\theta,\mathcal{R}) \triangleq \sqrt{\mathcal{R}^{2} - y^{2} \cos^{2} \left( \theta \right)} + y \sin \left( \theta \right)
\end{align}
\begin{align}
\mathscr{F}_{1} (z,p,\alpha,\mathcal{P},\mathcal{Q},T) & \triangleq \int^{T}_{0} \left(1-\left(1 + z p r^{-\alpha } \mathcal{Q} \right)^{-\mathcal{P}}\right) r \diff r \nonumber \\ & = \frac{1}{2} T^{2} \left(1-\frac{2}{\alpha \mathcal{P} +2} \left( \frac{T^{\alpha}}{z p \mathcal{Q}} \right)^{ \mathcal{P}} \, _2F_1\left(\mathcal{P},\mathcal{P}+\frac{2}{\alpha };\mathcal{P}+\frac{2}{\alpha }+1;-\frac{T^{\alpha }}{z p \mathcal{Q}}\right) \right)
\end{align}
\begin{align}
\mathscr{F}_{2} (z,p,\alpha,\mathcal{P},\mathcal{Q},T) & \triangleq \int^{+ \infty}_{T} \left(1-\left(1 + z p r^{-\alpha} \mathcal{Q} \right)^{-\mathcal{P}}\right) r \diff r = \frac{1}{2} T^2 \left(\, _2F_1\left(-\frac{2}{\alpha},\mathcal{P};1-\frac{2}{\alpha};- \frac{z p \mathcal{Q}}{T^{\alpha}} \right)-1\right).
\end{align}

\begin{thm}
\label{thm4}
In the FD C-RAN under disjoint clustering and capacity-limited fronthaul links, the achievable per-user DL and UL SEs (in nat/s/Hz) can be respectively upper-bounded by  	
\begin{multline}
\mathcal{S}^{(d)} \leq \int^{\mathcal{R}}_{0} \int^{C^{(d)}}_{0} \frac{1}{1 + x} \biggr( \frac{1}{2} + \frac{1}{\pi} \int^{+ \infty}_{0} \frac{1}{s} \operatorname{Im} \biggr( \mathcal{M}^{\text{d/j}}_{\mathcal{CMI}^{(d)}} \left( j s \right) \mathcal{M}^{\text{d/j}}_{\mathcal{ICI}^{(d)} | d} \left( j s \right) \\ \times \mathcal{M}^{\text{d/j}}_{\mathcal{X}^{(d)} | d} \left( \frac{- j s}{x} \right) \exp \left( j s \nu^{(d)}  \right) \biggr) \diff s \biggr) \frac{2 d}{\mathcal{R}^{2}} \diff x \diff d
\label{SE_DJ_DL}
\end{multline}
\begin{multline}
\mathcal{S}^{(u)} \leq \int^{\mathcal{R}}_{0} \int^{C^{(u)}}_{0} \frac{1}{1 + x} \biggr( \frac{1}{2} + \frac{1}{\pi} \int^{+ \infty}_{0} \frac{1}{s} \operatorname{Im} \biggr( 
 \mathcal{M}^{\text{d/j}}_{\mathcal{CMI}^{(u)}} \left( j s \right) \mathcal{M}^{\text{d/j}}_{\mathcal{ICI}^{(u)}} \left( j s \right) \\ \times  \mathcal{M}^{\text{d/j}}_{\mathcal{X}^{(u)} | d} \left( \frac{- j s}{x} \right) \exp \left( j s \nu^{(u)}  \right) \biggr) \diff s \biggr) \frac{2 d}{\mathcal{R}^{2}} \diff x \diff d
 \label{SE_DJ_UL}
\end{multline}
where
\begin{align}
\mathcal{M}^{\text{d/j}}_{\mathcal{X}^{(d)} | d}(j s) = \exp \left( - \lambda^{(d)} \int^{2 \pi}_{0} \mathscr{F}_{1} \left( j s,p^{(d)},\alpha,N^{(d)} - \mathcal{K}^{(d)} + \frac{1}{L},1,\Xi (d,\theta,\mathcal{R}) \right) \diff \theta \right)
\label{DL_SIGNAL_DJ}
\end{align}
\begin{align}
\mathcal{M}^{\text{d/j}}_{\mathcal{X}^{(u)} | d}(j s) = \exp \left( - \lambda^{(d)} \int^{2 \pi}_{0} \mathscr{F}_{1} \left( j s,p^{(u)},\alpha,N^{(u)} - \mathcal{K}^{(u)} + \frac{1}{L},1,\Xi (d,\theta,\mathcal{R}) \right) \diff \theta \right)
\label{UL_SIGNAL_DJ}
\end{align}
\begin{align}
\mathcal{M}^{\text{d/j}}_{\mathcal{ICI}^{(d)} | d} (j s) = \exp \left( - \lambda^{(d)} \int^{2 \pi}_{0} \mathscr{F}_{2} \left( j s,p^{(d)},\alpha,\mathcal{K}^{(d)},1,\Xi (d,\theta,\mathcal{R}) \right) \diff \theta \right)
\label{DL_ICI_DJ}
\end{align}
\begin{align}
\mathcal{M}^{\text{d/j}}_{\mathcal{ICI}^{(u)}} (j s) = \exp \left( - 2 \pi \lambda^{(d)} \int^{\mathcal{R}}_{0} \left( 1 - \exp \left( - \lambda^{(u)} \int^{2 \pi}_{0} \mathscr{F}_{2} \left( j s,p^{(u)},\alpha,\frac{1}{L},1,\Xi (y,\theta,\mathcal{R}) \right) \diff \theta \right) \right) y \diff y \right)
\label{UL_ICI_DJ}
\end{align} 
\begin{align}
\mathcal{M}^{\text{d/j}}_{\mathcal{CMI}^{(d)}} (j s) = \exp \left( - 2 \pi \lambda^{(u)} \mathscr{F}_{2} \left( j s,p^{(u)},\alpha,1,1,\mathcal{E} \right) \right)
\label{DL_CMI_DJ}
\end{align}
\begin{align}
\mathcal{M}^{\text{d/j}}_{\mathcal{CMI}^{(u)}} (j s) = \exp \left( - 2 \pi \lambda^{(d)} \int^{\mathcal{R}}_{0} \left( 1 - \exp \left( - \lambda^{(d)} \int^{2 \pi}_{0} \mathscr{F}_{2} \left( j s,p^{(d)},\alpha,\frac{\mathcal{K}^{(d)}}{L},1,\Xi (y,\theta,\mathcal{R}) \right) \diff \theta \right) \right) y \diff y \right). 
\label{UL_CMI_DJ}
\end{align}
Proof: {\normalfont See Appendix \ref{lem:proof}.}\end{thm}

\begin{thm}
\label{thm5}
In the FD C-RAN under user-centric clustering and capacity-limited fronthaul links, the achievable per-user DL and UL SEs (in nat/s/Hz) can be respectively upper-bounded by  
\begin{multline}
\mathcal{S}^{(d)} \leq \int^{C^{(d)}}_{0} \frac{1}{1 + x} \biggr( \frac{1}{2} + \frac{1}{\pi} \int^{+ \infty}_{0} \frac{1}{s} \operatorname{Im} \biggr( \mathcal{M}^{\text{u/c}}_{\mathcal{CMI}^{(d)}} \left( j s \right) \mathcal{M}^{\text{u/c}}_{\mathcal{ICI}^{(d)}} \left( j s \right) \mathcal{M}^{\text{u/c}}_{\mathcal{X}^{(d)}} \left( \frac{- j s}{x} \right) \exp \left( j s \nu^{(d)}  \right) \biggr) \diff s \biggr) \diff x
\label{SE_UC_DL}
\end{multline}
\begin{multline}
\mathcal{S}^{(u)} \leq \int^{C^{(u)}}_{0} \frac{1}{1 + x} \biggr( \frac{1}{2} + \frac{1}{\pi} \int^{+ \infty}_{0} \frac{1}{s} \operatorname{Im} \biggr( 
\mathcal{M}^{\text{u/c}}_{\mathcal{CMI}^{(u)}} \left( j s \right) \mathcal{M}^{\text{u/c}}_{\mathcal{ICI}^{(u)}} \left( j s \right) \mathcal{M}^{\text{u/c}}_{\mathcal{X}^{(u)}} \left( \frac{- j s}{x} \right) \exp \left( j s \nu^{(u)}  \right) \biggr) \diff s \biggr) \diff x
\label{SE_UC_UL}
\end{multline}
where
\begin{align}
\mathcal{M}^{\text{u/c}}_{\mathcal{X}^{(d)}}(j s) = \exp \left( - 2 \pi \lambda^{(d)} \mathscr{F}_{1} \left( j s,p^{(d)},\alpha,N^{(d)} - \mathcal{K}^{(d)} + \frac{1}{L},1,\mathcal{R} \right) \right)
\label{DL_SIGNAL_UC}
\end{align}
\begin{align}
\mathcal{M}^{\text{u/c}}_{\mathcal{X}^{(u)}}(j s) = \exp \left( - 2 \pi \lambda^{(d)} \mathscr{F}_{1} \left( j s,p^{(u)},\alpha,N^{(u)} - \mathcal{K}^{(u)} + \frac{1}{L},1,\mathcal{R} \right) \right)
\label{UL_SIGNAL_UC}
\end{align}
\begin{align}
\mathcal{M}^{\text{u/c}}_{\mathcal{ICI}^{(d)}} (j s) = \exp \left( - 2 \pi \lambda^{(d)} \mathscr{F}_{2} \left( j s,p^{(d)},\alpha,\mathcal{K}^{(d)},1,\mathcal{R} \right) \right)
\label{DL_ICI_UC}
\end{align}
\begin{align}
\mathcal{M}^{\text{u/c}}_{\mathcal{ICI}^{(u)}} (j s) = \exp \left( - 2 \pi \lambda^{(d)} \int^{\mathcal{R}}_{0} \left( 1 - \exp \left( - \lambda^{(u)} \int^{2 \pi}_{0} \mathscr{F}_{2} \left( j s,p^{(u)},\alpha,\frac{1}{L},1,\Xi (y,\theta,\mathcal{R}) \right) \diff \theta \right) \right) y \diff y \right)
\label{UL_ICI_UC}
\end{align} 
\begin{align}
\mathcal{M}^{\text{u/c}}_{\mathcal{CMI}^{(d)}} (j s) = \exp \left( - 2 \pi \lambda^{(u)} \mathscr{F}_{2} \left( j s,p^{(u)},\alpha,1,1,\mathcal{E} \right) \right)
\label{DL_CMI_UC}
\end{align}
\begin{align}
\mathcal{M}^{\text{u/c}}_{\mathcal{CMI}^{(u)}} (j s) = \exp \left( - 2 \pi \lambda^{(d)} \int^{\mathcal{R}}_{0} \left( 1 - \exp \left( - \lambda^{(d)} \int^{2 \pi}_{0} \mathscr{F}_{2} \left( j s,p^{(d)},\alpha,\frac{\mathcal{K}^{(d)}}{L},1,\Xi (y,\theta,\mathcal{R}) \right) \diff \theta \right) \right) y \diff y \right).
\label{UL_CMI_UC}
\end{align}
Proof: {\normalfont See Appendix \ref{lem:proof2}.}\end{thm}

\textit{Theorems 4} and \textit{5} provide complete solutions for the computation of the FD CRAN DL and UL SE upper-bounds under different disjoint and user-centric clustering approaches. In particular, the finite fronthaul capacities, $C^{(d)}$ and $C^{(u)}$, appear as limits of integration in the SE upper-bound expressions in (\ref{SE_DJ_DL}), (\ref{SE_DJ_UL}), (\ref{SE_UC_DL}), and (\ref{SE_UC_UL}) according to the cut-set theorem. Note that the exclusion region radius $\mathcal{E}$ in the UE-UE interference expression in (\ref{DL_CMI_DJ}) and (\ref{DL_CMI_UC}) can be tuned by design or measurements to capture the SIC capability at the reference UE. Further, the UL MI (i.e., BS-BS interference) is explicitly accounted for in (\ref{UL_CMI_DJ}) and (\ref{UL_CMI_UC}).

\section{Numerical Results}

In this section, we present some numerical examples in order to draw insights into the performance of FD versus HD C-RAN under different settings of system parameters. The density of the RUs is set to be $\lambda^{(d)} = \frac{4}{\pi}$ per km$^{2}$. The total system bandwidth is $W = 10$ MHz. The corresponding noise variance is set as $\nu^{(d)} = \nu^{(u)} = -174 + 10 \log_{10} ( W ) = - 104$ dBm. The DL (per user) and UL transmit powers are set as $23$ dBm and $20$ dBm, respectively. The results of the MC simulations are obtained based on $20000$ trials in a circular region of radius $50$ km. Recall that for the FD C-RAN, the DL and UL occur over the same time-frequency resources, where as in the HD C-RAN, the DL and UL are separated in the time domain. To facilitate comparison, we consider the per-user SE performance over two resource blocks. 

\subsection{Impact of Cooperation}

\begin{figure}
\centering
\begin{subfigure}{.5\textwidth}
\centering
\includegraphics[scale=1]{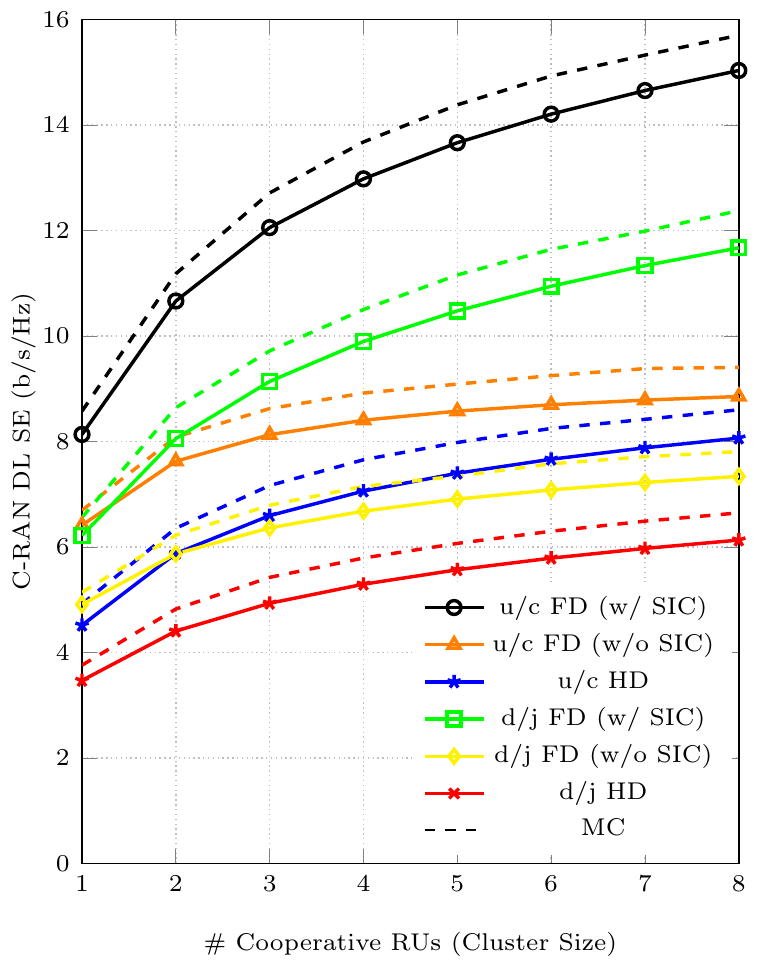}
\end{subfigure}%
\begin{subfigure}{.5\textwidth}
\centering
\includegraphics[scale=1]{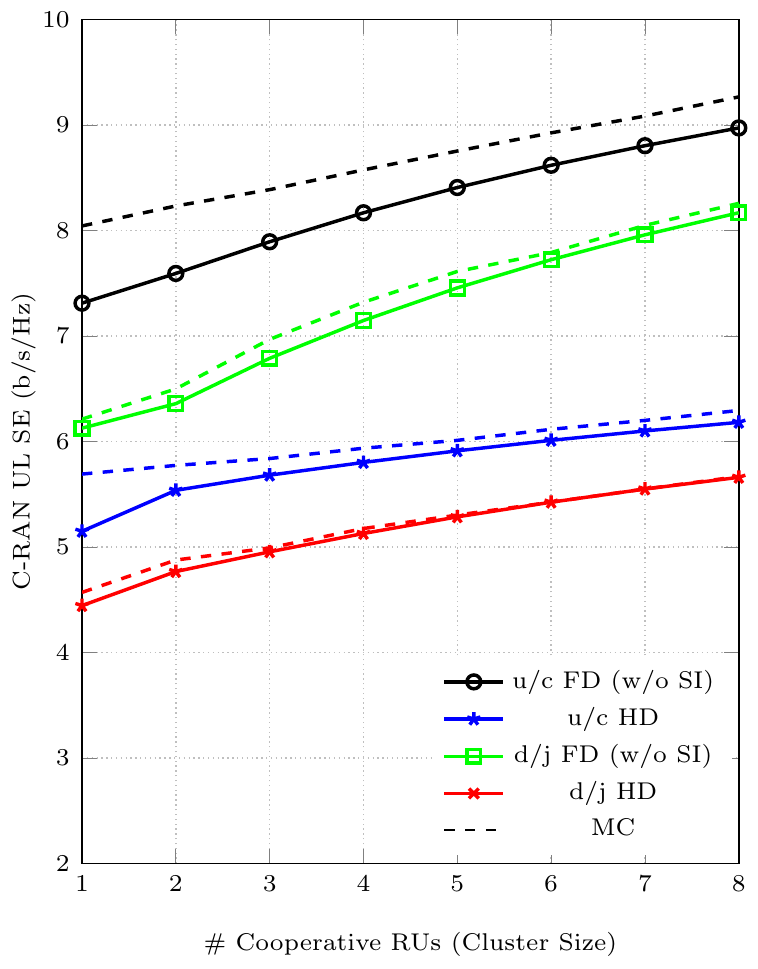}
\end{subfigure}
\caption{Impact of the number of cooperative RUs per cluster on the C-RAN SE performance. System parameters are: $\lambda^{(d)} = \frac{4}{\pi}$ RUs/km$^{2}$, $\lambda^{(u)} = \mathcal{K}^{(u)} \lambda^{(d)}$ UEs/km$^{2}$, $N^{(d)} = 8$, $N^{(u)} = 8$, $\mathcal{K}^{(d)} = 1$, $\mathcal{K}^{(u)} = 1$, $p^{(d)} = 0.2$ W, $p^{(u)} = 0.1$ W, $\nu^{(d)} = \nu^{(u)} = - 104$ dBm, $\alpha = 4$.} 
\label{FIG1}
\end{figure}

We investigate the impact of cooperation on the FD and HD C-RAN SEs under different clustering approaches and interference cancellation capabilities in Fig. \ref{FIG1}. In the DL, the SE always improves with larger cluster size ($L$). Furthermore, the FD over HD C-RAN DL SE gain (i) increases in $L$ with SIC capability (successful decoding and cancellation of the UL UE signal), and (ii) decreases in $L$ in the presence of intra-cluster MI. A similar trend can be observed in the UL, where the SE of the HD and FD C-RAN (with self-interference mitigation) improves as the number of cooperative RUs is increased. Here, the corresponding FD over HD C-RAN UL SE gain improves with increased cooperation. The highest FD versus HD C-RAN DL and UL SE gains recorded here are $86.4\%$ and $45.1\%$ (with $L = 8$), respectively. It is important to highlight that the SE performance of the user-centric clustering approach is superior to that of the disjoint method with relative greater improvement in the DL versus the UL. Note that the MC results confirm the validity of the proposed theoretical framework, with the gap in performance mostly stemming from the random per-cluster number of RUs in the MC simulations (versus the average number used in the theoretical analysis).     

\begin{figure}
\centering
\begin{subfigure}{.5\textwidth}
\centering
\includegraphics[scale=1]{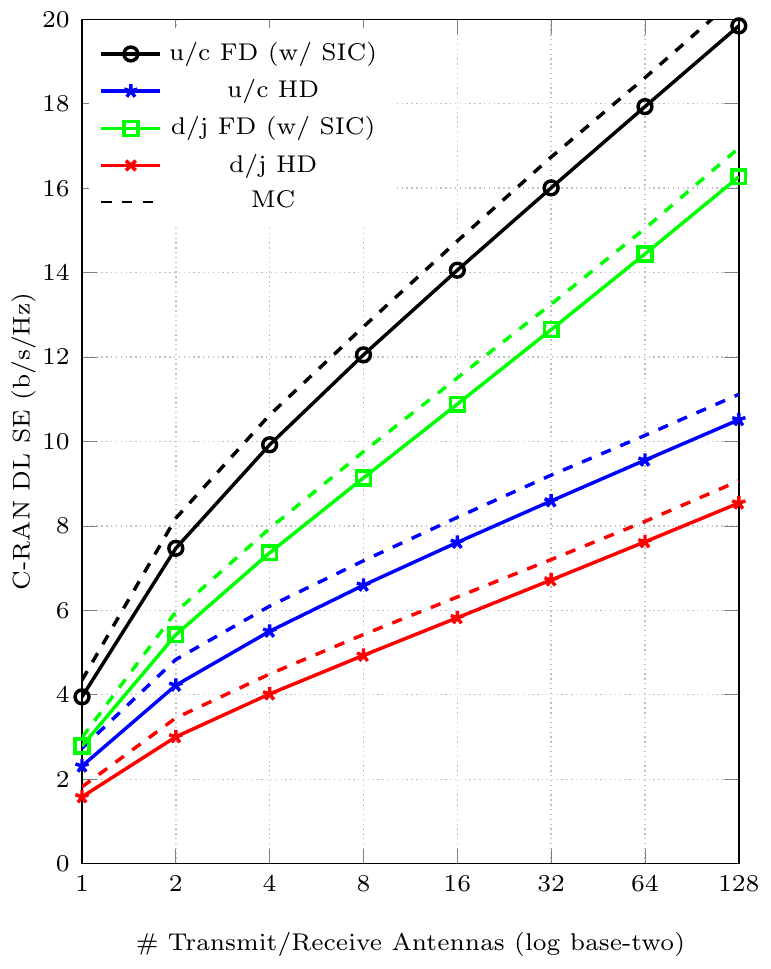}
\end{subfigure}%
\begin{subfigure}{.5\textwidth}
\centering
\includegraphics[scale=1]{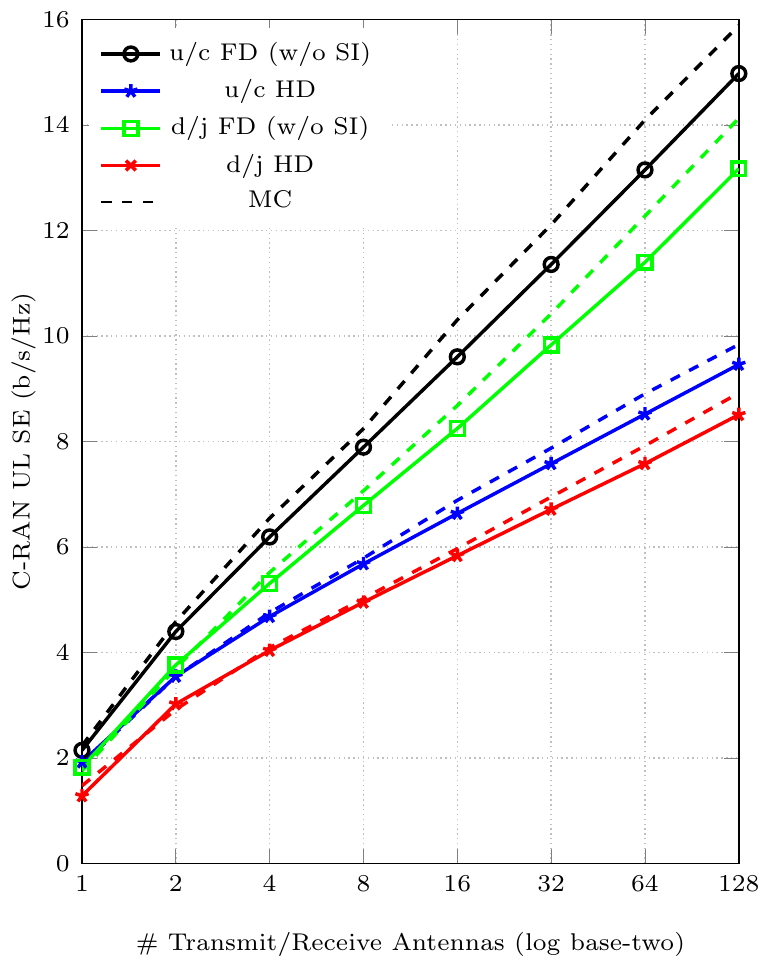}
\end{subfigure}
\caption{Impact of the RUs number of antennas on the C-RAN SE performance. System parameters are: $\lambda^{(d)} = \frac{4}{\pi}$ RUs/km$^{2}$, $\lambda^{(u)} = \mathcal{K}^{(u)} \lambda^{(d)}$ UEs/km$^{2}$, $L = 3$, $\mathcal{K}^{(d)} = 1$, $\mathcal{K}^{(u)} = 1$, $p^{(d)} = 0.2$ W, $p^{(u)} = 0.1$ W, $\nu^{(d)} = \nu^{(u)} = - 104$ dBm, $\alpha = 4$.}
\label{FIG2}
\end{figure}

\begin{figure}
\centering
\begin{subfigure}{.5\textwidth}
\centering
\includegraphics[scale=1]{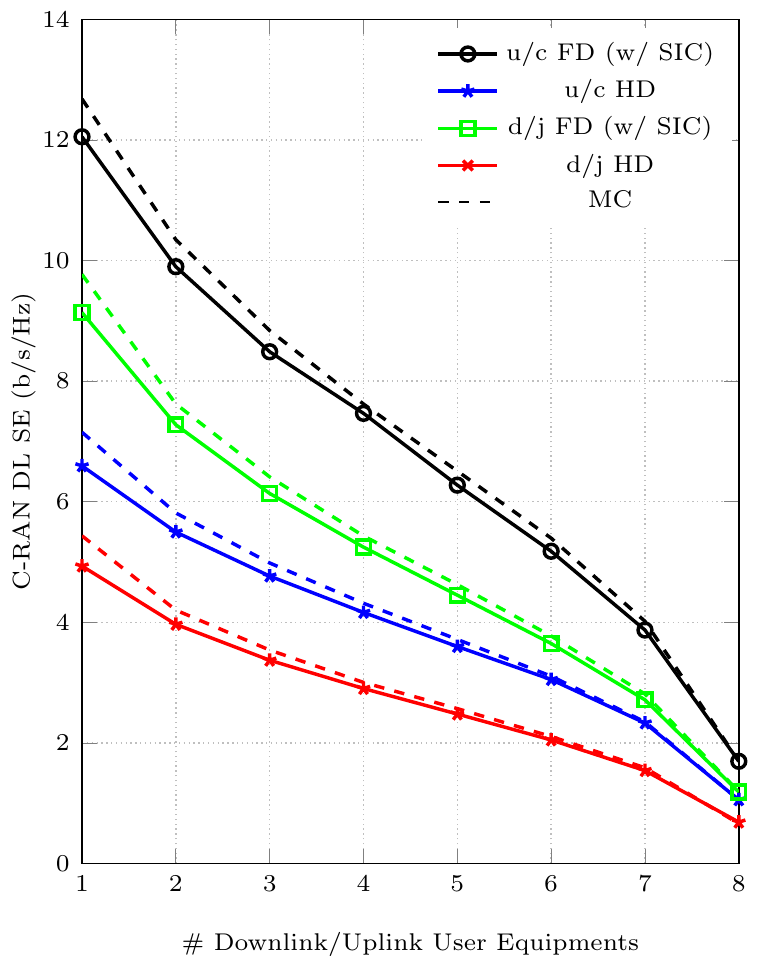}
\end{subfigure}%
\begin{subfigure}{.5\textwidth}
\centering
\includegraphics[scale=1]{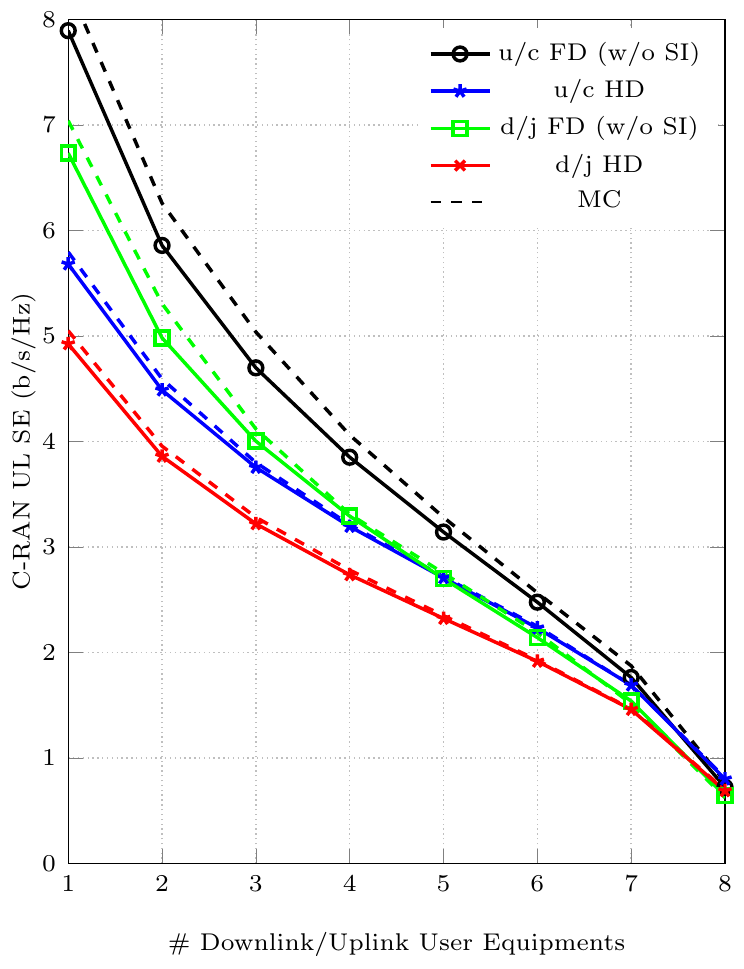}
\end{subfigure}
\caption{Impact of the number of UEs served by each RU per resource block on the C-RAN SE performance. System parameters are: $\lambda^{(d)} = \frac{4}{\pi}$ RUs/km$^{2}$, $\lambda^{(u)} = \mathcal{K}^{(u)} \lambda^{(d)}$ UEs/km$^{2}$, $L = 3$, $\mathcal{N}^{(d)} = 8$, $\mathcal{N}^{(u)} = 8$, $p^{(d)} = 0.2$ W, $p^{(u)} = 0.1$ W, $\nu^{(d)} = \nu^{(u)} = - 104$ dBm, $\alpha = 4$.}
\label{FIG3}
\end{figure}

\begin{figure}
\centering
\begin{subfigure}{.5\textwidth}
\centering
\includegraphics[scale=1]{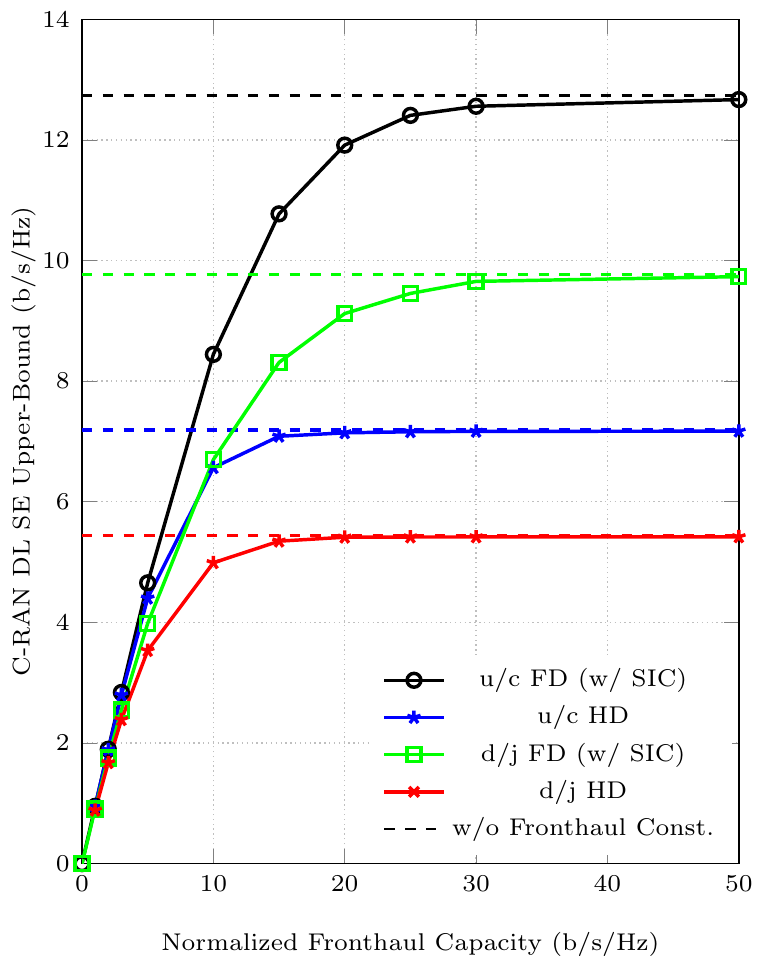}
\end{subfigure}%
\begin{subfigure}{.5\textwidth}
\centering
\includegraphics[scale=1]{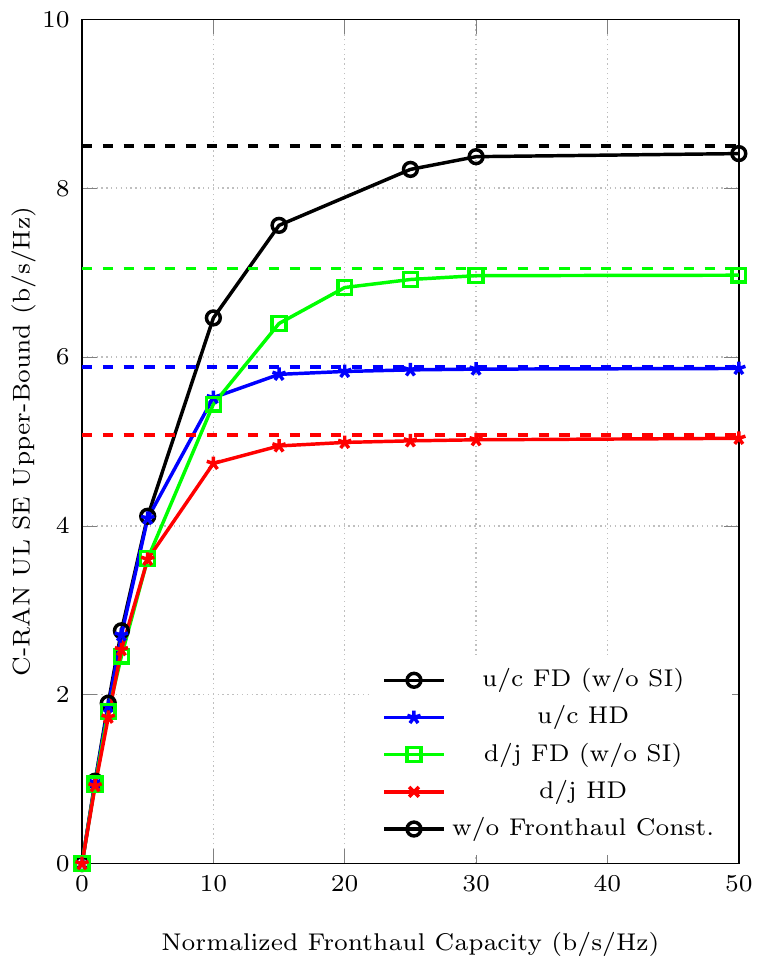}
\end{subfigure}
\caption{Impact of the fronthaul links normalized capacity on the C-RAN SE upper-bound performance. System parameters are: $\lambda^{(d)} = \frac{4}{\pi}$ RUs/km$^{2}$, $\lambda^{(u)} = \mathcal{K}^{(u)} \lambda^{(d)}$ UEs/km$^{2}$, $L = 3$, $N^{(d)} = 8$, $N^{(u)} = 8$, $\mathcal{K}^{(d)} = 1$, $\mathcal{K}^{(u)} = 1$, $p^{(d)} = 0.2$ W, $p^{(u)} = 0.1$ W, $\nu^{(d)} = \nu^{(u)} = - 104$ dBm, $\alpha = 4$.} 
\label{FIG4}
\end{figure}

\subsection{Impact of the Number of Antennas and User Equipments}

Next, we study the FD and HD C-RAN SE performances under different number of transmit/receive antennas at the RUs and number of DL/UL UEs. We capture performance under both disjoint and user-centric clustering approaches. In the case of FD C-RAN, we consider the case with SIC and self-interference cancellation capabilities. We can observe, based on the results from Fig. \ref{FIG2}, that with higher number of transmit/receive antennas at the RUs, a higher SE can be achieved. Furthermore, the FD over HD SE gain in both DL and UL directions of communications increases in the number of antennas at the RUs. On the other hand, the FD and HD C-RAN SE performance in the presence of different number of DL/UL users is shown in Fig. \ref{FIG3}. It can be observed that the (per-user) SE performance degrades in the number of active users. Furthermore, the FD over HD SE gain in both the DL and the UL is reduced as we increase the number of users served per resource block. 

\subsection{Impact of Fronthaul}

We investigate the impact of capacity-limited fronthaul on the FD and HD C-RAN performance in Fig. \ref{FIG4}. In the case of HD C-RAN, the fronthaul links are dedicated to either the DL or the UL operation per resource block, whereas in the case of FD C-RAN, the fronthaul capacity is divided equally between the DL and UL operations per resource block. As expected, increasing the fronthaul capacity enhances the C-RAN SE upper-bound performance. Furthermore, the FD over HD C-RAN SE upper-bound gain in both DL and UL increases with greater fronthaul capacity. The capacity of fronthaul links in practice are anticipated to be one or more orders of magnitude greater than the DL or the UL SE \cite{6845310}. The results from Fig. \ref{FIG4} illustrate that in such cases both FD and HD SE values converge, hence, with sufficient-capacity fronthaul, significant SE gains can be achieved through the FD operation at the RU side. It is important to note that the SE curves under the fronthaul constraint capture the upper-bound performance based on the cut-set theorem. In practice, the achievable SE, depending on the relaying strategy, is further impacted by other factors arising from the finite-capacity fronthaul links (such as compression noise). A rigorous investigation of this aspect is left for future work.

\section{Conclusions}

This paper provided a stochastic design and analysis of large-scale C-RAN with FD enabled RUs. Different disjoint and user-centric approaches for the formation of the finite clusters were considered. We incorporated the notion of non-isotropic fading channels in characterizing the distribution of the signals power gains. Upper-bound expressions of SE were accordingly derived, in particular, accounting for the impact of finite-capacity fronthaul links and MI between the DL and the UL. The results indicated that the FD over HD SE gain can be enhanced considerably through cooperative communications capabilities of C-RAN. The corresponding improvement in SE performance is particularly evident with advanced interference cancellation strategies and sufficient-capacity fronthaul links.  

\appendices
\numberwithin{equation}{section}

\section{}
\label{apx:proof}

The DL and UL intended channels strengths in the FD C-RAN can be respectively expressed as $\| \boldsymbol{g}_{c,c_{o}} \|^{2} = \sum_{c_{j} \in \Phi^{(d)}_{c}} \beta_{c_{j},c_{o}}$ $\| \boldsymbol{f}_{c_{j},c_{o}} \|^{2}$ and $\| \boldsymbol{h}_{c_{i},c} \|^{2} = \sum_{c_{l} \in \Phi^{(d)}_{c}} \beta_{c_{i},c_{l}} \| \boldsymbol{f}_{c_{i},c_{l}} \|^{2}$. Through applying the Gamma moment matching technique \cite{7450690}, we can respectively obtain $\| \boldsymbol{g}_{c,c_{o}} \|^{2} \approx \mathpzc{G} ( \kappa_{c,c_{o}} , \theta_{c,c_{o}})$ and $\| \boldsymbol{h}_{c_{i},c} \|^{2} \approx \mathpzc{G} (\kappa_{c_{i},c} , \theta_{c_{i},c})$ where $\kappa_{c,c_{o}} \triangleq N^{(d)} ( \sum_{c_{j} \in \Phi^{(d)}_{c}} \beta_{c_{j},c_{o}} )^2 / ( \sum_{c_{j} \in \Phi^{(d)}_{c}}$\linebreak $\beta^{2}_{c_{j},c_{o}} )$, $\theta_{c,c_{o}} = ( \sum_{c_{j} \in \Phi^{(d)}_{c}} \beta^{2}_{c_{j},c_{o}} ) / (\sum_{c_{j} \in \Phi^{(d)}_{c}} \beta_{c_{j},c_{o}} )$, $\kappa_{c_{i},c} \triangleq N^{(u)} ( \sum_{c_{l} \in \Phi^{(d)}_{c}} \beta_{c_{i},c_{l}} )^{2} / ( \sum_{c_{l} \in \Phi^{(d)}_{c}} \beta^{2}_{c_{i},c_{l}} )$, and $\theta_{c_{i},c} \triangleq ( \sum_{c_{l} \in \Phi^{(d)}_{c}}$\linebreak $\beta^{2}_{c_{i},c_{l}} ) / ( \sum_{c_{l} \in \Phi^{(d)}_{c}} \beta_{c_{i},c_{l}} )$. Next, by invoking the approach from \cite{7268913}, we assume that the DL and UL intended channels are isotropic with approximate distributions $\boldsymbol{g}_{c,c_{o}} \approx \mathpzc{G} (\boldsymbol{0}, \theta_{c,c_{o}} \mathbf{I}_{L_{c} N^{(t)}})$ and $\boldsymbol{h}_{c_{i},c} \approx \mathpzc{G} (\boldsymbol{0}, \theta_{c_{i},c} \mathbf{I}_{L_{c} N^{(r)}})$, respectively. It can be shown that the DL and UL cooperative ZF beamforming spaces in the cluster $c$ are respectively $L_{c} ( N^{(d)} - \mathcal{K}^{(d)} ) + 1$ and $L_{c} ( N^{(u)} - \mathcal{K}^{(u)} ) + 1$ dimensional \cite{1261332}. Hence, when projecting the intended DL and UL channels onto the cooperative ZF precoding and decoding subspaces, we consider each spatial dimension (i.e., antenna) respectively adds $\frac{\kappa_{c,c_{o}}}{L_{c} N^{(d)}} (L_{c} ( N^{(d)} - \mathcal{K}^{(d)} ) + 1)$ and $\frac{\kappa_{c_{i},c}}{L_{c} N^{(u)}} (L_{c} ( N^{(u)} - \mathcal{K}^{(u)} ) + 1)$ to the corresponding Gamma distributed channel power gains. Hence, we can approximate the DL and UL channel power gains using equivalent distributions $| \boldsymbol{g}_{c,c_{o}} \boldsymbol{v}_{c,c_{o}} |^{2} \approx \sum_{c_{j} \in \Phi^{(d)}_{c}} \beta_{c_{j},c_{o}} \hat{\psi}_{c_{j},c_{o}}, \;\, \hat{\psi}_{c_{j},c_{o}} \sim \mathpzc{G} (  N^{(d)} - \mathcal{K}^{(d)} + \tfrac{1}{L_{c}} , 1 )$ and $| \boldsymbol{w}^{T}_{c_{i},c} \boldsymbol{h}_{c_{i},c} |^{2} \approx \sum_{c_{l} \in \Phi^{(d)}_{c}} \beta_{c_{i},c_{l}} \hat{\psi}_{c_{i},c_{l}}, \;\, \psi_{c_{i},c_{l}} \sim \mathpzc{G} ( N^{(u)} - \mathcal{K}^{(u)} + \tfrac{1}{L_{c}} , 1 )$, respectively. To facilitate stochastic performance analysis, we further approximate the DL and UL intended channel power gains based on the average number of cooperating RUs in each cluster, $L$. 

The channel power gain expression for the DL inter-cluster interference is derived using a similar approach to that described above. We characterize the DL inter-cluster interference channel strength using Gamma moment matching. Then, the corresponding channel, approximated via an isotropic distribution, is projected onto the one-dimensional interference vector space. Furthermore, under the assumption of inter-cluster precoding matrices having independent column vectors \cite{7478073}, we approximate the respective channel power gain using the Gamma distribution, with each aggregate inter-cluster channel contributing $\mathcal{K}^{(d)}$ to the shape parameter. The channel power gain of the UL inter-cluster interference can be readily derived using the same approach, with each link between the UL UE interferer to a cooperating RU in the reference cluster contributing $\frac{1}{L}$ to the corresponding Gamma distributed channel power gain shape parameter.

The cross-mode channels between the UL and DL UEs in the FD C-RAN are isotropic in nature, as there is no coordination among the UEs. Hence, we can readily express the corresponding channel power gains by separating the small- and large-scale fading effects as in (\ref{eq:DLcmi}). On the other hand, the cross-mode channels between the RUs, involve the squared norm of a vector with each element being a sum of non-identically distributed random variables, i.e., in the form $\| \boldsymbol{w} \boldsymbol{G} \boldsymbol{V} \|^2$, where $\boldsymbol{w}$, $\boldsymbol{G}$, and $\boldsymbol{V}$ denote the decoding vector, MIMO fading channel, and precoding matrix, respectively. In a recent contribution in \cite{7805138}, we provided a unified approximate expression for the distribution of $\| \boldsymbol{w} \boldsymbol{F} \boldsymbol{V} \|^2$, considering arbitrary linear beamforming design, and isotropic MIMO channels. By invoking the assumption that the precoding matrices have independent columns, and using a similar moment matching technique described previously, we can approximate the cross-mode interference channel power gain between two clusters. The corresponding Gamma distribution is from the aggregation of the power of each link between a RU (from an interfering cluster) transmit antennas to a RU receive antennas (in the reference cluster). Hence, we can arrive at the expression in (\ref{eq:ULcmi}).

\section{}
\label{thm1:proof}

Consider $\mathbb{E} \left\{ \min \left( \log (1 + \gamma) , C \right) \right\}$, where $C$ is a non-negative constant constraint (e.g., backhaul or fronthaul SE) on the achievable capacity. This expectation can be equivalently expressed using the complimentary CDF of the SINR as
\begin{align}
\mathbb{E} \left\{ \min \left( \log (1 + \gamma) , C \right) \right\} & = \int^{+ \infty}_{0} \mathscr{P} ( \min \left( \log (1 + \gamma) , C \right) > \tau ) \diff \tau \nonumber \\ & = \int^{+ \infty}_{0} \mathscr{P} ( \log (1 + \gamma) > \tau , C  > \tau ) \diff \tau \nonumber \\ & = \int^{C}_{0} \mathscr{P} ( \log (1 + \gamma) > \tau ) \diff \tau \nonumber \\ & = \int^{C}_{0} \frac{1}{1+x} \mathscr{P} ( \gamma > x ) \diff x. 
\end{align}
Hence, we arrive at the results in \textit{Theorem \ref{thm1}}. 
\hfill $\blacksquare$

\section{}
\label{thm2:proof}

Consider the general SINR expression $\gamma = \frac{X}{\sum_{i} I_{i} + \nu}$. The CDF of $\gamma$ can hence be formulated as
\begin{align}
\mathcal{F}_{\gamma} (x) = \mathscr{P} \left( \gamma \leq x \right) = \mathscr{P} \left( \sum_{i} I_{i} - \frac{X}{x} \geq - \nu \right) \overset{(i)}{=} \mathscr{P} \left( \Theta \geq - \nu \right) 
\end{align}
where $\Theta \triangleq \sum_{i} I_{i} - \frac{X}{x}$. By applying the Gil-Pelaez inversion theorem \cite{GilPelaez}
\begin{align}
\mathscr{P} \left( \Theta \geq - \nu \right) = \frac{1}{2} - \frac{1}{\pi} \int^{+ \infty}_{0} \frac{1}{s} \operatorname{Im} \left( \mathcal{M}_{\Theta} \left( j s \right) \exp \left( j s \nu \right) \right) \diff s.
\end{align}
Using the above result, and with the MGF of $\Theta$, in the case that $X$, $\forall I_{i}$ are independent, we arrive at \textit{Theorem \ref{thm2}}. \hfill $\blacksquare$

\section{}
\label{lem:proof}

In the case of disjoint clustering, a typical user location is a random variable following a uniform distribution in the corresponding cluster area. The statistics of the DL intended signal in the FD C-RAN under disjoint clustering can be derived as
\begin{multline}
\mathcal{M}^{\text{d/j}}_{\mathcal{X}^{(d)}}(z) = \mathbb{E} \left\{ \exp \left( - z \mathcal{X}^{(d)} \right) \right\} = \mathbb{E} \left\{ \exp \left( - z p^{(d)} \left| \boldsymbol{g}_{c,c_{o}} \boldsymbol{v}_{c,c_{o}} \right|^{2}  \right) \right\} \\ \overset{(i)}{\approx} \mathbb{E} \left\{ \exp \left( - z p^{(d)} \sum_{c_{j} \in \Phi^{(d)}_{c}} \beta_{c_{j},c_{o}} \psi_{c_{j},c_{o}} \right) \right\} \overset{(ii)}{=} \mathbb{E}_{\Phi^{(d)}_{c}} \left\{ \prod_{c_{j} \in \Phi^{(d)}_{c}} \mathbb{E}_{\psi_{c_{j},c_{o}}} \left\{ \exp \left( - z p^{(d)} \beta_{c_{j},c_{o}} \psi_{c_{j},c_{o}} \right) \right\} \right\} \\ \overset{(iii)}{=} \mathbb{E}_{\Phi^{(d)}_{c}} \left\{ \prod_{c_{j} \in \Phi^{(d)}_{c}} \left( 1 + z p^{(d)} \beta_{c_{j},c_{o}} \right)^{- \left( N^{(d)} - \mathcal{K}^{(d)} + \frac{1}{L} \right)} \right\} \\ \overset{(iv)}{=} \exp \left( - \lambda^{(d)} \int^{2 \pi}_{0} \int^{\Xi (y,\theta) }_{0} \left( 1 - \left( 1 + z p^{(d)} r^{- \alpha} \right)^{- \left( N^{(d)} - \mathcal{K}^{(d)} + \frac{1}{L} \right)}  \right) r \diff r \diff \theta\right) 
\end{multline}
where $(i)$ follows from \textit{Approximation 1}; $(ii)$ is from the independence property of PPP and uncorrelated channel conditions; $(iii)$ is obtained through the MGF a Gamma random variable, i.e., for $X \sim \mathpzc{G} (\mathcal{P},\mathcal{Q})$, we have $\mathbb{E}_{X} \left\{ \exp (- z X) \right\} = (1 + z \mathcal{Q})^{-\mathcal{P}}$; $(iv)$ is written using the probability generating functional (PGFL), i.e., for a stationary PPP $\Phi$ with density $\lambda$, we have $\mathbb{E}_{\Phi} \left\{ \prod_{\boldsymbol{x} \in \Phi} f (\boldsymbol{x}) \right\} = \exp \left( - \lambda \int_{\mathbb{R}^{2}} \left( 1 - f \left( \boldsymbol{x} \right) \right) \diff \boldsymbol{x} \right)$, and converting from Cartesian to polar coordinates (with Jacobian $r$), where considering a circular cluster of radius $\mathcal{R}$, the distance between a typical RU-UE pair with an angle $\theta$ can range from zero to $\Xi (y,\theta) \triangleq \sqrt{\mathcal{R}^{2} - y^{2} \cos^{2} \left( \theta \right)} + y \sin \left( \theta \right)$; and finally, we arrive at (\ref{DL_SIGNAL_DJ}) by adopting the integral identity $\mathscr{F}_{1} (z,p,\alpha,\mathcal{P},\mathcal{Q},T) \triangleq \int^{T}_{0} \left(1-\left(1 + z p r^{-\alpha } \mathcal{Q} \right)^{-\mathcal{P}}\right) r \diff r = \frac{1}{2} T^{2} \left(1-\frac{2}{\alpha \mathcal{P} +2} \left( \frac{T^{\alpha}}{z p \mathcal{Q}} \right)^{ \mathcal{P}} \, _2F_1\left(\mathcal{P},\mathcal{P}+\frac{2}{\alpha };\mathcal{P}+\frac{2}{\alpha }+1;-\frac{T^{\alpha }}{z p \mathcal{Q}}\right) \right)$. Using the same methodology, the statistics of the UL intended signal is given by
\begin{align}
\mathcal{M}^{\text{d/j}}_{\mathcal{X}^{(u)}}(z) = \mathbb{E} \left\{ \exp \left( - z \mathcal{X}^{(u)} \right) \right\} = \mathbb{E} \left\{ \exp \left( - z p^{(u)} \left| \boldsymbol{w}^{T}_{c_{i},c} \boldsymbol{h}_{c_{i},c} \right|^{2} \right) \right\} = (\ref{UL_SIGNAL_DJ}). 
\end{align}

Moreover, the DL ICI in the FD C-RAN under disjoint clustering is given by
\begin{multline}
\mathcal{M}^{\text{d/j}}_{\mathcal{ICI}^{(d)}} (z) = \mathbb{E} \left\{ \exp \left( - z \mathcal{ICI}^{(d)} \right) \right\} = \mathbb{E} \left\{ \exp \left( - z p^{(d)} \sum_{m \in \Psi \setminus\{ c \}} \left\| \boldsymbol{g}_{m,c_{o}} \boldsymbol{V}_{m} \right\|^2 \right) \right\} \\ \overset{(i)}{\approx} \mathbb{E} \left\{ \exp \left( - z p^{(d)} \sum_{m \in \Psi \setminus\{ c \}} \sum_{m_{j} \in \Phi^{(d)}_{m}} \beta_{m_{j},c_{o}} \psi_{m_{j},c_{o}}  \right) \right\} \overset{(ii)}{=} \mathbb{E} \left\{ \exp \left( - z p^{(d)} \sum_{b \in \Phi^{(d)} \setminus \mathcal{B}(\mathcal{R})} \beta_{b,c_{o}} \psi_{b,c_{o}}  \right) \right\} \\ = \mathbb{E}_{\Phi^{(d)}} \left\{ \prod_{b \in \Phi^{(d)} \setminus \mathcal{B}(\mathcal{R})} \mathbb{E}_{\psi_{b,c_{o}}} \left\{ \exp \left( - z p^{(d)} \beta_{b,c_{o}} \psi_{b,c_{o}}  \right) \right\} \right\} = \mathbb{E}_{\Phi^{(d)}} \left\{ \prod_{b \in \Phi^{(d)} \setminus \mathcal{B}(\mathcal{R})} \left( 1 + z p^{(d)} \beta_{b,c_{o}} \right)^{- \mathcal{K}^{(d)}} \right\} \\ = \exp \left( - \lambda^{(d)} \int^{2 \pi}_{0} \int^{+ \infty}_{\Xi (y,\theta)} \left( 1 - \left( 1 + z p^{(d)} r^{- \alpha} \right)^{- \mathcal{K}^{(d)}} \right) r \diff r \diff \theta \right) 
\end{multline}
where $(i)$ holds under \textit{Approximation 2}; $(ii)$ follows from the aggregate interference from inter-cluster RUs being equivalent in distribution to the total interference generated by PPP-based RUs that are independently transmitting outside the circular ball ($\mathcal{B}$) of radius $\mathcal{R}$; and (\ref{DL_ICI_DJ}) is obtained by using the integral identity $\mathscr{F}_{2} (z,p,\alpha,\mathcal{P},\mathcal{Q},T) \triangleq \int^{+ \infty}_{T} \left(1-\left(1 + z p r^{-\alpha} \mathcal{Q} \right)^{-\mathcal{P}}\right) r \diff r = \frac{1}{2} T^2 \left(\, _2F_1\left(-\frac{2}{\alpha},\mathcal{P};1-\frac{2}{\alpha};- \frac{z p \mathcal{Q}}{T^{\alpha}} \right)-1\right)$.

The UL ICI statistics in the FD C-RAN under disjoint clustering is given by
\begin{multline}
\mathcal{M}^{\text{d/j}}_{\mathcal{ICI}^{(u)}} (z) = \mathbb{E} \left\{ \exp \left( - z \mathcal{ICI}^{(u)} \right) \right\} = \mathbb{E} \left\{ \exp \left( - z p^{(u)} \sum_{m \in \Psi \setminus \{ c \} } \sum_{m_{k} \in \Psi^{(u)}_{m}} \left| \boldsymbol{w}^{T}_{c_{i},c} \boldsymbol{h}_{m_{k},c} \right|^{2} \right) \right\} \\ \overset{(i)}{\approx} \mathbb{E} \left\{ \exp \left( - z p^{(u)} \sum_{m \in \Psi \setminus \{ c \},m_{k} \in \Psi^{(u)}_{m}} \sum_{c_{l} \in \Phi^{(d)}_{c}} \beta_{m_{k},c_{l}} \psi_{m_{k},c_{l}} \right) \right\} \\ \overset{(ii)}{=} \mathbb{E} \left\{ \exp \left( - z p^{(u)} \sum_{c_{l} \in \Phi^{(d)}_{c}} \sum_{k \in \Phi^{(u)} \setminus \mathcal{B} (\mathcal{R})}  \beta_{k,c_{l}} \psi_{k,c_{l}} \right) \right\} \\ \overset{(iii)}{=} \mathbb{E}_{\Phi^{(d)}_{c}} \left\{ \prod_{c_{l} \in \Phi^{(d)}_{c}} \mathbb{E}_{\Phi^{(u)}} \left\{ \prod_{k \in \Phi^{(u)} \setminus \mathcal{B} (\mathcal{R})} \mathbb{E}_{\psi_{k,c_{l}}} \left\{ \exp \left( - z p^{(u)} \beta_{k,c_{l}} \psi_{k,c_{l}} \right) \right\} \right\} \right\} \\ \overset{(iv)}{=} \mathbb{E}_{\Phi^{(d)}_{c}} \left\{ \prod_{c_{l} \in \Phi^{(d)}_{c}} \mathbb{E}_{\Phi^{(u)}} \left\{ \prod_{k \in \Phi^{(u)} \setminus \mathcal{B} (\mathcal{R})} \left( 1 + z p^{(u)} \beta_{k,c_{l}} \right)^{- \frac{1}{L}} \right\} \right\} \\ \overset{(v)}{=} \mathbb{E}_{\Phi^{(d)}_{c}} \left\{ \prod_{c_{l} \in \Phi^{(d)}_{c}} \exp \left( - \lambda^{(u)} \int^{2 \pi}_{0} \int^{+ \infty}_{\Xi (y , \theta)} \left( 1 - \left( 1 + z p^{(u)} r^{- \alpha} \right)^{- \frac{1}{L}} \right) r \diff r \diff \theta \right) \right\} \\ \overset{(vi)}{=} \exp \left( - 2 \pi \lambda^{(d)} \int^{\mathcal{R}}_{0} \left( 1 - \exp \left( - \lambda^{(u)} \int^{2 \pi}_{0} \int^{+ \infty}_{\Xi (y , \theta)} \left( 1 - \left( 1 + z p^{(u)} r^{- \alpha} \right)^{- \frac{1}{L}} \right) r \diff r \diff \theta \right) \right) y \diff y \right) 
\end{multline} 
where $(i)$ follows from \textit{Approximation 2}; $(ii)$ is written considering equivalence in distribution; $(iii)$ is based on the independence property of PPP and uncorrelated channel conditions; $(iv)$ is from the MGF of a Gamma random variable; $(v)$ follows from applying PGFL with respect to the PPP of UEs; $(vi)$ follows from applying PGFL with respect to the PPP of RUs; and (\ref{UL_ICI_DJ}) is obtained using the integral identity $\mathscr{F}_{2}(.)$.

Next, under disjoint clustering, we characterize the DL cross-mode interference considering the UEs may be capable of performing SIC. In order to capture performance for general cases, we consider an exclusion region of radius $\mathcal{E}$ when modeling the DL cross-mode interference. Hence, we can obtain 
\begin{multline}
\mathcal{M}^{\text{d/j}}_{\mathcal{CMI}^{(d)}} (z) = \mathbb{E} \left\{ \exp \left( - z  \mathcal{CMI}^{(d)} \right) \right\} = \mathbb{E} \left\{ \exp \left( - z p^{(u)} \sum_{m \in \Psi,m_{k} \in \Psi^{(u)}_{m}} \left| h_{m_{k},c_{o}} \right|^2 \right) \right\} \\ \overset{(i)}{\approx} \mathbb{E} \left\{ \exp \left( - z p^{(u)} \sum_{k \in \Phi^{(u)} \setminus \mathcal{B} (\mathcal{E})} \left| h_{k,c_{o}} \right|^2 \right) \right\} \overset{(ii)}{=} \mathbb{E}_{\Phi^{(u)}} \left\{ \prod_{k \in \Phi^{(u)} \setminus \mathcal{B} (\mathcal{E})} \mathbb{E}_{\psi_{k,c_{o}}} \left\{ \exp \left( - z p^{(u)} \beta_{k,c_{o}} \psi_{k,c_{o}} \right) \right\} \right\} \\ \overset{(iii)}{=} \mathbb{E}_{\Phi^{(u)}} \left\{ \prod_{k \in \Phi^{(u)}} \left( 1 + z p^{(u)} \beta_{k,c_{o}} \right)^{-1} \right\} \overset{(iv)}{=} \exp \left( - 2 \pi \lambda^{(u)} \int^{\infty}_{\mathcal{E}} \left( 1 - \left( 1 + z p^{(u)} r^{- \alpha}_{k,c_{o}} \right)^{-1} \right) r \diff r \right) 
\end{multline}
where $(i)$ is written based on \textit{Approximation 3}; $(ii)$ follows from the independence property of PPP and uncorrelated channel conditions; $(iii)$ is from the MGF of a Gamma random variable; $(iv)$ holds via applying PGFL; and (\ref{DL_CMI_DJ}) is written using the integral identity $\mathscr{F}_{2} (.)$. 

On the other hand, the statistics of the UL cross-mode interference in the FD C-RAN can be characterized as 
\begin{multline}
\mathcal{M}^{\text{d/j}}_{\mathcal{CMI}^{(u)}} (z) = \mathbb{E} \left\{ \exp \left( - z  \mathcal{CMI}^{(u)} \right) \right\} = \mathbb{E} \left\{ \exp \left( - z p^{(d)} \sum_{m \in \Psi \setminus \{ c \}} \left\| \boldsymbol{w}^{T}_{c_{i},c} \boldsymbol{G}_{m,c} \boldsymbol{V}_{m} \right\|^{2} \right) \right\} \\ \overset{(i)}{\approx} \mathbb{E} \left\{ \exp \left( - z p^{(d)} \sum_{m \in \Psi \setminus \{ c \},m_{j} \in \Phi^{(d)}_{m}} \sum_{c_{l} \in \Phi^{(d)}_{c}} \beta_{m_{j},c_{l}} \psi_{m_{j},c_{l}} \right) \right\} \\ \overset{(ii)}{=} \mathbb{E} \left\{ \exp \left( - z p^{(d)} \sum_{c_{l} \in \Phi^{(d)}_{c}} \sum_{b \in \Phi^{(d)} \setminus \mathcal{B}(\mathcal{R})} \beta_{b,c_{l}} \psi_{b,c_{l}} \right) \right\} \\ \overset{(iii)}{=} \mathbb{E}_{\Phi^{(d)}_{c}} \left\{ \prod_{c_{l} \in \Phi^{(d)}_{c}} \mathbb{E}_{\Phi^{(d)}} \left\{ \prod_{b \in \Phi^{(d)} \setminus \mathcal{B}(\mathcal{R})} \mathbb{E}_{\psi_{b,c_{l}}} \left\{ \exp \left( - z p^{(d)} \beta_{b,c_{l}} \psi_{b,c_{l}} \right) \right\} \right\} \right\} \\ \overset{(iv)}{=} \mathbb{E}_{\Phi^{(d)}_{c}} \left\{ \prod_{c_{l} \in \Phi^{(d)}_{c}} \mathbb{E}_{\Phi^{(u)}} \left\{ \prod_{b \in \Phi^{(d)} \setminus \mathcal{B}(\mathcal{R})} \left( 1 + z p^{(d)} \beta_{b,c_{l}} \right)^{- \frac{\mathcal{K}^{(d)}}{L}} \right\} \right\} \\ \overset{(v)}{=} \mathbb{E}_{\Phi^{(d)}_{c}} \left\{ \prod_{c_{l} \in \Phi^{(d)}_{c}} \exp \left( - \lambda^{(d)} \int^{2 \pi}_{0} \int^{+ \infty}_{\Xi \left( y,\theta \right)} \left( 1 - \left( 1 + z p^{(d)} r^{- \alpha} \right)^{- \frac{\mathcal{K}^{(d)}}{L}} \right) r \diff r \diff \theta \right) \right\} 
\end{multline}
where $(i)$ is written based on \textit{Approximation 3}; $(ii)$ follows from equivalence in distribution; $(iii)$ holds considering the independence property of PPP and uncorrelated channel conditions; $(iv)$ is written using the MGF of a Gamma random variable; $(v)$ follows by taking the PGFL with respect to the inter-cluster RUs; $(vi)$ is obtained through the PGFL with respect to the intra-cluster RUs; and we arrive at (\ref{UL_CMI_DJ}) using the integral identity $\mathscr{F}_{1}(.)$.
 
Hence, we arrive at \textit{Theorem \ref{thm4}}. \hfill $\blacksquare$
 
\section{Proof of the Signals Statistics Expressions}
\label{lem:proof2} 
 
In the case of user-centric clustering, each UE is always at the center of its formed cluster. Hence, the statistics of the different intended signals can be readily derived as
\begin{multline}
\mathcal{M}^{\text{u/c}}_{\mathcal{X}^{(d)}}(z) = \mathbb{E} \left\{ \exp \left( - z \mathcal{X}^{(d)} \right) \right\} = \mathbb{E} \left\{ \exp \left( - z p^{(d)} \left| \boldsymbol{g}_{c,c_{o}} \boldsymbol{v}_{c,c_{o}} \right|^{2}  \right) \right\} \\ \approx \exp \left( - 2 \pi \lambda^{(d)} \int^{\mathcal{R}}_{0} \left( 1 - \left( 1 + z p^{(d)} r^{- \alpha} \right)^{- \left( N^{(d)} - \mathcal{K}^{(d)} + \frac{1}{L} \right)}  \right) r \diff r \right) = (\ref{DL_SIGNAL_UC})
\end{multline}
\begin{multline}
\mathcal{M}^{\text{u/c}}_{\mathcal{X}^{(u)}}(z) = \mathbb{E} \left\{ \exp \left( - z \mathcal{X}^{(u)} \right) \right\} = \mathbb{E} \left\{ \exp \left( - z p^{(u)} \left| \boldsymbol{w}^{T}_{c_{i},c} \boldsymbol{h}_{c_{i},c} \right|^{2} \right) \right\} \\ \approx \exp \left( - 2 \pi \lambda^{(d)} \int^{\mathcal{R}}_{0} \left( 1 - \left( 1 + z p^{(u)} r^{- \alpha} \right)^{- \left( N^{(u)} - \mathcal{K}^{(u)} + \frac{1}{L} \right)}  \right) r \diff r \right) = (\ref{UL_SIGNAL_UC}).
\end{multline}

Moreover, the DL ICI in the FD C-RAN under user-centric clustering can be obtained as
\begin{multline}
\mathcal{M}^{\text{u/c}}_{\mathcal{ICI}^{(d)}} (z) = \mathbb{E} \left\{ \exp \left( - z \mathcal{ICI}^{(d)} \right) \right\} = \mathbb{E} \left\{ \exp \left( - z p^{(d)} \sum_{m \in \Psi \setminus\{ c \}} \left\| \boldsymbol{g}_{m,c_{o}} \boldsymbol{V}_{m} \right\|^2 \right) \right\} \\ \approx \exp \left( - 2 \pi \lambda^{(d)} \int^{+ \infty}_{\mathcal{R}} \left( 1 - \left( 1 + z p^{(d)} r^{- \alpha} \right)^{- \mathcal{K}^{(d)}} \right) r \diff r \right) = (\ref{DL_ICI_UC}).
\end{multline}
The above should be viewed as an upper-bound approximation given that under user-centric clustering, the cooperative RUs may be serving different UEs, hence, there may exist certain intra-cluster interference. 

On the other hand, it can be shown that the DL cross-mode interference as well as the different UL interference terms are equivalent in distribution under disjoint and user-centric clustering approaches. Hence, we have (\ref{UL_ICI_UC}), (\ref{DL_CMI_UC}), and (\ref{UL_CMI_UC}). 

Hence, we arrive at \textit{Theorem \ref{thm5}}. \hfill $\blacksquare$

\bibliographystyle{IEEEtran}
\bibliography{IEEEabrv,myref}

\end{document}